\documentclass[twocolumn, 10pt,letterpaper]{IEEEtran}
\usepackage[font=small]{caption}
\usepackage{etex}
\usepackage{graphicx}
\usepackage{tikz}
\usetikzlibrary{shapes,arrows}
\usetikzlibrary{arrows}
\usepackage{amsfonts} 
\usepackage{tabularx}
\usepackage{subfig}
\usepackage{amssymb}
\usepackage{booktabs}
\usepackage{caption}
\usepackage{capt-of}
\usepackage{multirow}
\usepackage{paralist}
\usepackage{subfloat}
\usepackage{array}
\usepackage{verbatim}
\usepackage{multicol}
\usepackage{multirow}
\usepackage{amsthm}
\usepackage{booktabs}
\usepackage{varwidth}
\usepackage{algorithm}
\usepackage[noend]{algpseudocode}
\theoremstyle{definition}

\newcolumntype{L}{>{\centering\arraybackslash}m{2.5cm}}
\newcolumntype{M}{>{\centering\arraybackslash}m{1cm}}
\newcolumntype{P}{>{\centering\arraybackslash}m{2cm}}
\newcolumntype{N}{>{\centering\arraybackslash}m{3cm}}
\newcolumntype{Q}{>{\centering\arraybackslash}m{4cm}}
\usepackage{afterpage}
\usepackage[nodisplayskipstretch]{setspace}

\usepackage{epsfig}
\usepackage{epstopdf}
\usepackage{array}

\usepackage[noadjust]{cite}

\usepackage{atbegshi}
\AtBeginDocument{\AtBeginShipoutNext{\AtBeginShipoutDiscard}}

\usepackage[cmex10]{amsmath}
\allowdisplaybreaks
\usepackage[utf8]{inputenc}

\newcommand{\Kc}{{\cal K}}
\newcommand{\wv}{{\mathbf w}}
\newcommand{\sv}{{\mathbf s}}

\newcommand{\Sc}{\mathcal{S}} 
\newcommand{\PP}{\mathbb{P}}
\newcommand{\CaRReM}{CAREM}

\begin{document}
\title{A Context-aware Radio Resource Management\\ in Heterogeneous Virtual RANs}
\author{Sharda Tripathi, Corrado Puligheddu, Carla Fabiana Chiasserini, Federico Mungari} 
\thanks{S. Tripathi is with Birla Institute of Technology and Science-Pilani, Pilani Campus, India, C.  Puligheddu, C. F. Chiasserini, F. Mungari are with Politecnico di Torino, Italy (e-mail: sharda.tripathi@pilani.bits-pilani.ac.in, \{corrado.puligheddu,~carla.chiasserini,~federico.mungari\}@polito.it). C. F. Chiasserini is also with CNIT, Italy and CNR-IEIIT, Italy.
This work has been supported by the EC H2020 5GPPP  5GROWTH project (Grant No.\,856709).

© 2021 IEEE. Personal use of this material is permitted. Permission
from IEEE must be obtained for all other uses, in any current or future
media, including reprinting/republishing this material for advertising or
promotional purposes, creating new collective works, for resale or
redistribution to servers or lists, or reuse of any copyrighted
component of this work in other works.
}

\maketitle
\pagenumbering{arabic}

\begin{abstract}
New-generation wireless networks are designed to support a wide range of services with diverse key performance indicators (KPIs) requirements. A fundamental component of such networks, and a pivotal factor to the fulfillment of the target KPIs, is the virtual radio access network (vRAN), which allows high flexibility on the control of the radio link. However, to fully exploit the potentiality of vRANs, an efficient mapping of the rapidly varying context to radio control decisions is not only essential, but also challenging owing to the interdependence of user traffic demand, channel conditions, and resource allocation.
Here, we propose CAREM, a reinforcement learning framework for dynamic radio resource allocation in heterogeneous vRANs, which selects the best available link and transmission parameters for packet transfer, so as to meet the KPI requirements.
To show its effectiveness, we develop a testbed for proof-of-concept. Experimental results demonstrate that CAREM enables an efficient radio resource allocation under different settings and traffic demand. Also, compared to the closest existing scheme based on neural network and the standard LTE, CAREM exhibits an improvement of one order of magnitude in packet loss and latency, while it provides a 65\% latency improvement relatively to the contextual bandit approach.
\end{abstract}


\section{Introduction}
The envisaged paradigm of new-generation mobile technologies is aimed to serve a broad spectrum of applications having diverse requirements on various key performance indicators (KPIs), ranging from high reliability and low latency to large-scale connectivity and massive data rates \cite{andrews}. To accommodate such ambitious vision, new generation wireless access networks are required  not only to integrate various flexible multi-access technologies such as mmWave and massive MIMO \cite{wang}, but also to provide a versatile radio resource management (RRM) system that can ensure efficient spectrum utilization and seamless interoperability \cite{olwal}. 

A  powerful concept addressing such needs is the  virtualization of  the radio access network (RAN),  
wherein the legacy communication system is decoupled by centralizing the softwarized radio access through virtual machines or containers running on servers  at the edge of the cellular network \cite{liang2015,tsagkaris2015,tsagkaris2015}. While this makes the network more agile and minimizes the requirement of expensive dedicated hardware, the edge may host several applications competing for resources, thereby limiting the efficiency of radio functions \cite{romero}. Besides, with an unprecedented increase in the number of devices trying to concurrently access the virtual RAN (vRAN), 
it is expected to observe in the near future a 1,000-fold growth in network traffic \cite{5G}. This will contribute to complex interference dynamics  and will require sophisticated techniques for RRM, which can effectively cope with both  the  diverse performance requirements of the applications to be supported and the rapidly varying network and channel conditions.

Evidently, the unification of hybrid technologies under the new-generation cellular umbrella adds to the complexity of the problem, thereby making
the use of conventional  theoretic approaches often inadequate to achieve optimum traffic and resource management,  owing to intricate mathematical modeling and complex dependencies between network and channel variables. It has thus become  indispensable to design innovative solutions that can  effectively deal with  the system complexity thanks to a fully automated, data-driven approach. 

Recently, machine learning (ML) techniques have shown to hold an enormous potential in addressing the challenges of applying standard mathematical optimization frameworks to resource allocation problems in vRANs and in allowing an automatic system control \cite{fu2018}. A plethora of learning-based techniques including supervised, unsupervised, reinforcement learning (RL), and deep learning have been proposed \cite{hussain2020,tang2020,xiong2019} for heterogeneous networks in general, to tackle resource allocation problems (see Sec.\,\ref{related_works} for a more detailed discussion). However, it is worth noting that, while deep learning approaches are computationally intensive, the primary challenge associated with simpler ones such as supervised/unsupervised learning is the creation of an exhaustive dataset for training the model. Besides, in case of a rapidly changing environment, frequent retraining of the model is required to achieve the desired accuracy, which can be expensive when there are stringent latency constraints. To this end, it is required to devise a framework that is easy to train in non-stationary environments, yet effective in making intelligent choices in an autonomous fashion using near real-time feedback on channel conditions and temporal variation of user demand so as to improve performance and reliability of the network. 

In this work, we leverage the advantages offered by ML and develop a context-aware, RL-based solution to radio resource management  in heterogeneous vRANs.
Our scheme, named \CaRReM\ (Context-Aware Radio rEsource Management), is devised considering 
a formulation for RRM based on sequential decision making \cite{zhengEbook}, which, thanks to a persistent interaction between the learning agent and its environment, can effectively cope with time-varying operating conditions. 
The key contributions of this work are as follows:
\begin{enumerate}

\item We design \CaRReM , a   framework using differential semi-gradient State-Action-Reward-State-Action (SARSA)  for periodic RRM in a multi-user  vRAN scenario. \CaRReM\ efficiently 
identifies the radio link to be used, allocates radio resources, and sets such transmission parameters for packet transfer as the modulation and coding scheme (MCS) while meeting two of the main KPI requirements identified by 3GPP \cite{3gpp}, namely, packet loss and  latency.

\item Since each heterogeneous link features a maximum available resource capacity, we define an algorithm ensuring that, if multiple users  are assigned to the same link, the allocation is Pareto-efficient fair and the overall allocated resources  do not exceed the link maximum capacity.

\item We investigate the complexity of the proposed \CaRReM\ framework, and introduce a two-fold approach to expedite the convergence. Firstly, high dimensionality of context variables is addressed using a practical tile coding approach. Secondly, the action space is designed as a subset of discrete positive integers, which in turn limits its cardinality and facilitates simultaneous selection of several action components (e.g., link, fraction of radio resources, MCS) using a single action.   

\item A proof-of-concept is provided in the context of 
heterogeneous communications and multiple users, by designing a testbed  implementing \CaRReM\ over 3GPP LTE and IEEE 802.11p links using software defined radios (SDR).

\item The \CaRReM\ performance is evaluated under different settings, including different decision periodicity, number of links and connected users, and values of traffic load.   The results show that, as the learning converges, \CaRReM\ can be efficiently used for link, MCS and radio resource selection in vRANs. With respect to the 100-ms decision, 1-s decisions significantly reduce the computational demand, without any noticeable performance degradation.  Further, when compared against the closest existing RRM technique  in \cite{romero}, contextual bandit approach \cite{slivkins}, and standard LTE, CAREM always shows significantly better performance, with  an improvement of one order of magnitude in both packet loss and latency with respect to the radio policy in \cite{romero} and standard LTE, and a 65\% latency improvement relatively to contextual bandit.

\end{enumerate}

We remark that the use of a reward signal for associating the best decisions to different contexts, and the dependence of future contexts on current decisions are two important aspects of this problem which makes it different from a much simpler contextual bandit formulation \cite{slivkins}. To effectively tackle these challenges, we adopt a model-free, full-blown, RL approach using the differential semi-gradient SARSA algorithm. Unlike Q-learning \cite{jang}, which is a popular off-policy RL approach  useful for episodic tasks, SARSA has low per-sample variance, thereby making it less susceptible to convergence problems. Also, in a continuous task setting such as RRM where it is required to care for agent's performance during the exploration phase, online learning using SARSA is preferred due to its conservative nature of avoiding high risk actions that generate large negative rewards from the environment. To  our knowledge, no existing work has presented such comprehensive and dynamic framework for RRM, keen on fast and reliable data transmission in heterogeneous vRANs.

The rest of the paper is organized as follows. Sec.\,\ref{related_works} reviews the related works, while   Sec.\,\ref{system_model} and Sec.\,\ref{learning_framework} introduce, respectively, our system model  and the proposed RL framework. Sec.\,\ref{testbed_design} describes the implementation of our solution and the developed testbed, and Sec.\,\ref{results} discusses performance evaluation results. Finally,  Sec.\,\ref{conclusion} concludes the paper.

\section{Related Work} \label{related_works}
Owing to the intricate channel-network dynamics and complexity of  heterogeneous networks, several works have aimed at devising strategies for effective resource utilization, while meeting the stringent KPI requirements of different applications to be supported in a cellular network. The state-of-the-art primarily revolves around the idea of learning environment variables and their evolution over time for optimizing resource utilization and improving real-time system performance. In particular,  learning-based techniques have been developed to address multichannel access, scheduling and allocation of resource blocks, modulation and coding schemes, computation resources, transmit power and data rate, while  maximizing KPI satisfaction, with particular emphasis on throughput, latency, packet loss, channel utility, and user fairness. 

The problems of dynamic rate allocation as well as of joint channel and rate selection for throughput maximization have been studied in \cite{combes2015, combes2019, gupta2018, ma2019, hasegawa2020, qureshi2020 }. While \cite{combes2015, combes2019} use a multi-armed bandit formulation and exploit unimodal feature of reward over the arms using UCB policies, an algorithm based on Thompson sampling is used in \cite{gupta2018} for achieving link-rate selection in logarithmic time regret. Likewise, a ML approach, also based on multi-armed bandit using tug-of-war dynamics, is presented in \cite{ma2019, hasegawa2020} for channel selection in IoT networks. These works however do not explore the contextual information from the environment for transmission parameter optimization. To address this limitation and further improving the performance, a structured RL approach using contextual unimodal multi-armed bandit is proposed in \cite{qureshi2020}, for dynamic rate selection and distributed resource allocation.

Unlike the bandit model, the RL approach is more popular in recent literature for radio resource provisioning problems, especially if the action corresponds to a decision making scenario with discrete choices. RL-based schemes are proposed for selecting the radio access technology  in heterogeneous networks using network-centric \cite{helou2015}, and user-centric approaches \cite{nguyen2017}. 
In \cite{wei2018}, a policy gradient actor-critic algorithm is studied for user scheduling and resource allocation in energy-efficient heterogeneous networks. The works in \cite{morozs2015} and \cite{raj2018} investigate dynamic spectrum access in cognitive radio networks using the RL framework, with the aim  to achieve high controllability in spectrum sharing and to minimize the sensing duration. 
Owing to delay-sensitivity and massive volume of data traffic in 5G access networks, an RL-based scheduling scheme is introduced in \cite{comsa2018, comsa2020} to minimize the packet delay and drop rate. The study in \cite{romero}, instead, proposes a deep deterministic policy gradient algorithm based on actor-critic neural network and a classifier  for resource control decisions. This is the most relevant work to ours, as it specifically addresses a virtualized access network and presents an implementation of the solution in a full-fledged testbed. Under high mobility and high traffic demand, RL-based radio resource control in 5G vehicular networks is tackled  in \cite{zhou2020}, with the goal of  adaptively changing uplink to downlink ratio in a frequency band.

Advanced ML such as deep learning techniques are of interest for resource allocation problems when the size of state-action space is large, leading to slow convergence of RL approaches. A deep Q-network for channel selection is proposed in \cite{wang2018, naparstek2019} to adaptively learn in time-varying scenarios subject to maximization of network utility. The study in \cite{zong2019} envisions an adaptive deep actor-critic, RL-based framework for channel access in dynamic environment for multi-user scenarios. Deep RL is explored for selection of suitable MCS for primary transmissions in cognitive radio networks in \cite{zhang2019}. 
In a similar setting, the study in \cite{li2018} investigates a deep learning dynamic power control method for a secondary user to coexist with the primary user. A distributed dynamic power allocation using multi-agent deep RL is developed in \cite{nasir2019}, which exploits  channel state and quality of service information to maximize a sum-rate utility function. Deep RL is also applicable to radio resource management in vehicular networks including channel selection, optimal sub-band allocation, and power control, as shown in  \cite{gyawali2019, chen2020, ye2019, zhang2019_iot}. 

At last, we mention that a preliminary version of this work has appeared in our conference paper \cite{wons2021}.

{\em Novelty.} 
First, unlike most of prior art on RRM,  we address the selection of link, radio resources, and packet transmission parameters,  so that  the target values of packet loss rate and latency KPIs are achieved for each traffic flow. 
While  in terms of actions of the RL framework  \cite{nguyen2017,zhang2019_iot,zhang2019} are somewhat aligned to our work, their learning objectives and KPI requirements are very different. 
Furthermore, with respect to all the above works, including \cite{romero}, we address connectivity between a radio point of access and multiple users over {\em heterogeneous} links. In addition, the RRM policy in \CaRReM\ (i) is spontaneously learned and updated over time by its continuous interaction with the environment, thus being able to adapt continuously to time-varying channel and network dynamics, and (ii)  provides a fair Pareto-efficient allocation of capacity-constrained resources, thus leading to an effective management of multi-user connectivity.  
Finally, we provide a proof-of-concept of the proposed solution, implementing it in a multi-technology, multi-user, SDR-based testbed.

\begin{figure}[!t]
\centering{{\includegraphics[width=0.9\linewidth]{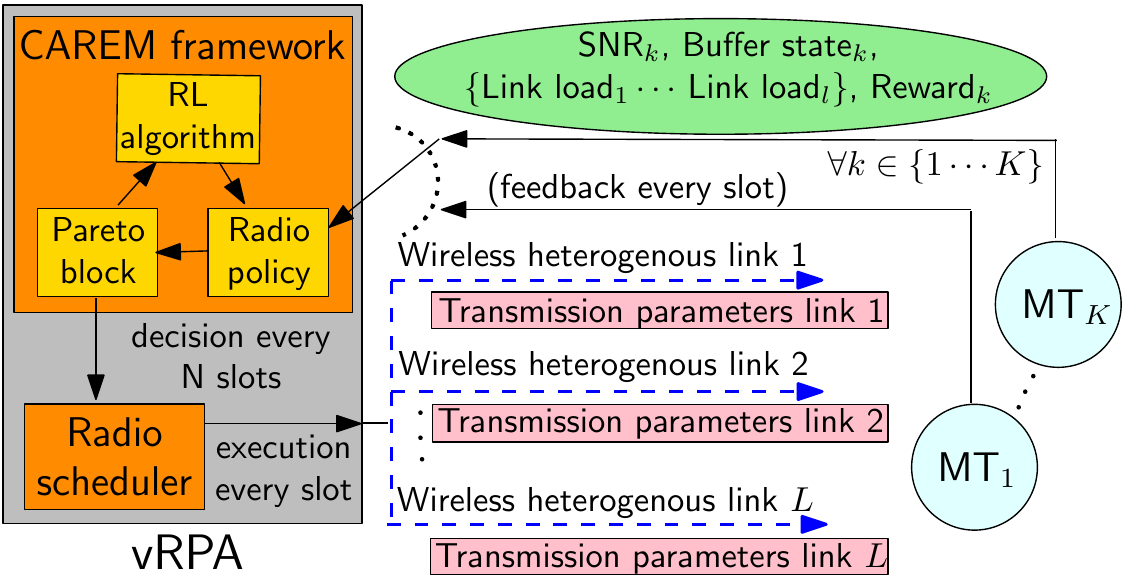} }}
   \caption{vRAN system model and architecture of \CaRReM\ framework.
   { \CaRReM\ gathers per slot contextual information for each active MT and makes decisions every $N$ slots. Through an RL algorithm, a radio policy and a Pareto block, it selects the best available radio link and transmission parameters for every MT, which is input to the radio scheduler that executes these decisions every slot.}
}
\label{fig:fig_sysmodel}
\vspace{-3mm}
  \end{figure}

  \section{System Architecture} \label{system_model}
In this section, we present the system model considered for provisioning of radio resources via \CaRReM. 
Although our approach and methodology are general and can apply to any number and type of vRAN technologies, while describing the framework  we refer for concreteness to  the  communication environment we implemented in our testbed where LTE and IEEE 802.11p links are available. 

We leverage SDR interfaces enabling 
point-to-point communications between virtual radio point of access (vRPA) and $K$ users, hereinafter referred to as  Mobile Terminals (MTs), 
implemented at the edge of the network. The architecture of the proposed \CaRReM\ framework in a vRAN is presented in Fig.\,\ref{fig:fig_sysmodel}.
We also envision that user applications such as video streaming, gaming, road safety services (e.g., for vehicles or vulnerable road users)  are deployed at the edge through containerized infrastructure, and possibly co-located with 
radio functions including radio resource management, scheduling, admission control, and reliable packet delivery. 
In the following, we focus on the downlink data transfers from the vRPA to the MTs, although our framework can be easily extended to uplink scenarios as well. 

On the cellular downlink, the vRPA determines the number of radio resources per MT, i.e., of resource blocks (RBs),  required for the transmission of data packets, based upon the signal-to-noise ratio (SNR) reported by each MT through the Channel Quality Indicator (CQI). Conversely, on the IEEE 802.11p link, the vRPA accesses the channel to transmit to the MTs using the CSMA-based scheme  foreseen by the corresponding standard 
To minimize packet loss over  the radio links, at the physical layer  data packets are modulated and encoded using a suitable MCS (namely, twenty-nine and eight different MCS values are possible on the LTE and IEEE 802.11p links, respectively). 
Furthermore, at the MAC layer,  an automatic repeat request error control is in place, i.e., an unsuccessfully transmitted packet can be  resent till a maximum number of allowed retransmission attempts. Beside SNR, the knowledge of the amount of data waiting to be transmitted towards a MT and of the traffic load supported on available links is also essential for radio resources provisioning, so as to minimize packet loss. The information on the buffer occupancy can be acquired using buffer state reports at the MAC layer. 

The proposed \CaRReM\ framework is a dynamic resource controller that is included as an extended functionality within the  vRPA and that interacts with  the  radio scheduler implemented therein. 
At the logical level, it is composed of  as many  RL agents as the number of quality of Service (QoS) classes to be supported, each agent differing from the others in the target KPIs. 
Then,  each RL agent can sequentially handle multiple MTs. The RL agent  considers the status corresponding to each MT, namely, SNR, buffer state, and also the status of aggregate traffic load already hosted on  the available  links, and selects the  best action. The latter includes: link, MCS value, and number of RBs or channel utilization time, so as to maximize the associated reward, hence meet the  KPI  requirements. 

\begin{figure}[bt]
\centering{\includegraphics[width=0.9\linewidth]{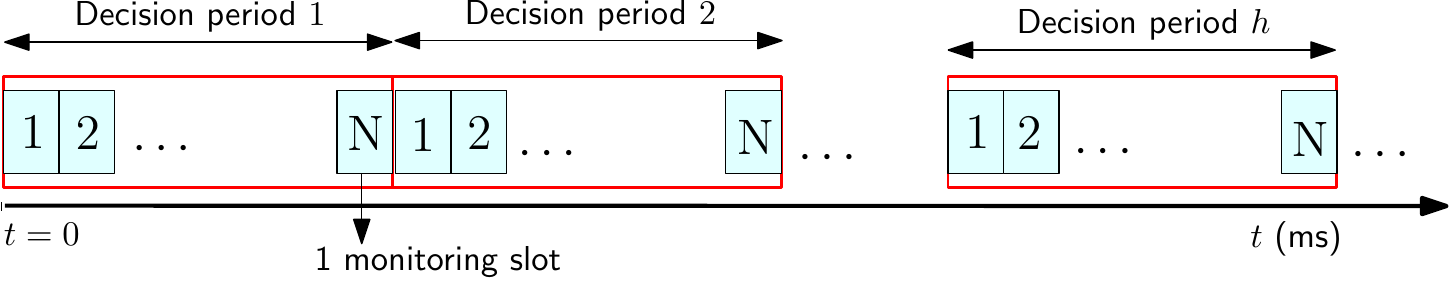}}
   \caption{Relation between a decision period and monitoring slots, {each decision period consists of $N$ monitoring slots.}}
\label{fig:DP_MS_relation}
\vspace{-3mm}
  \end{figure}

For each MT, the SNR, buffer state, and link load values are periodically monitored in a time-slotted fashion. 
From the learning of the environment variables, a decision on the action to be adopted 
is made by the CAREM framework every $N$ time slots. During the decision making process, the link load values are continuously updated as the RL agent sequentially selects an action for each MT.
Subsequently, these decisions are enforced as radio policies by the scheduler on a per-slot basis, till the next decision-making event.
Specifically, while the periodicity with which the scheduler operates remains constant, we consider that CAREM  may make decisions with periodicity equal to  $N\geq 1$ slots.  A clear demarcation of decision period and monitoring slot is depicted in Fig.\,\ref{fig:DP_MS_relation}.

\section{The \CaRReM\ Framework} \label{learning_framework}

The joint impact of channel and network dynamics on RRM in wireless networks is far from being trivial; therefore, 
for efficient resource mapping in non-stationary environments, we adopt a model-free approach that does not require an environment model. 
Our CAREM framework, depicted in Fig.\,\ref{fig:fig_sysmodel}, continuously maps the variations in transmission channel and traffic load into a context, thus learning  the best action for each given context. 
For sake of clarity and without loss of generality, below we focus on a single RL agent, handling multiple MTs connected to the vRPA and receiving traffic flows belonging to the same QoS class. 

\begin{figure}[!t]
\centering{{\includegraphics[width=0.5\linewidth]{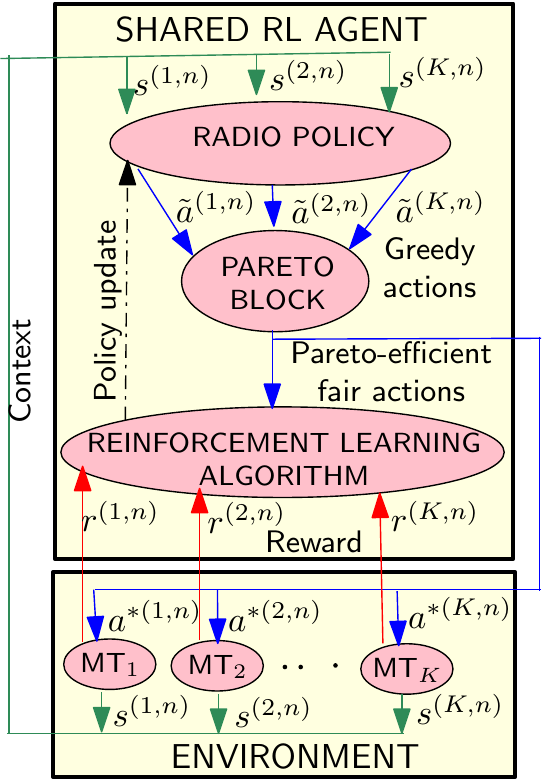} }}
   \caption{Components of the \CaRReM\ framework. { Radio policy maps contextual information gathered from the environment into actions. Whenever needed, the Pareto block refines the actions so that they meet the link capacity. The RL algorithm updates the radio policy using differential semi-gradient SARSA.}}
\label{fig:fig_rlcomp}
\vspace{-3mm}
  \end{figure}

Each RL agent includes three blocks, interacting with each other as depicted in Fig.\,\ref{fig:fig_rlcomp}: 
(i) the radio policy selecting  the best action for each MT connection, 
(ii) the Pareto block, refining the previous action with respect to the allocation of capacity-constrained resources, and
(iii) the RL algorithm implementing a differential semi-gradient SARSA.
Note that in all our experiments we observed that a single-agent implementation can cope with a large number of MTs, while entailing a negligible latency due to the
decision-making process. Nevertheless, whenever the container, or virtual machine, in
which the RL agent is implemented needs to be scaled out due to an exceedingly high computing
burden, such techniques as those developed, e.g., within the  5G-Transformer \cite{5GT} and the 5Growth \cite{5Growth} projects, can be effectively applied.

Below, we  detail the three main components of  CAREM.

\subsection {Radio policy}  
 The radio policy block continuously maps observation of contexts from the environment to decisions in the form of actions, for each MT. 
 The goal of the RL model is to train the agent to find a policy that eventually maximizes the cumulative reward from an uncertain environment.
The elements composing the radio policy  are introduced below; unless otherwise specified, we refer to a generic decision-making period, which is composed of $N$ monitoring slots.

{\bf Context Space.}  For the generic MT $k$ ($k \in \{1,\ldots, K\}$) connected to the vRPA, in monitoring slot $n$ ($n=1,\ldots,N$) the agent observes a context vector $s^{(k,n)} \in \mathcal{X} $,
applies  action $a^{(k)} \in \mathcal{A}$,  which was selected at the end of the previous decision period and holds for the whole current one, and receives a reward value $r(s^{(k,n)}, a^{(k)})$ as feedback. 
As discussed in Sec.\,\ref{system_model}, the environment variables, namely, SNR, buffer state, and links load, influence the choice of the link, MCS, and radio resource allocation. Let  $\gamma^{(k,n)}$ and $\sigma^{(k,n)}$ be, respectively, the SNR and the buffer state reported by the $k$-th MT during the $n$-th monitoring slot. Also, let $\zeta^{(k,l,n)}$ denote the aggregate link load already on  link $l$ ($l \in \{1, \ldots, L\}$) during the $n$-th monitoring slot while making a decision for MT $k$.  
We can then write the generic context vector as $s^{(k,n)} := \{\gamma^{(k,n)}, \sigma^{(k,n)}, \zeta^{(k,1,n)} \cdots \zeta^{(k,L,n)} \}$.

{\textbf {Action Space.}} 
Let us denote the amount of capacity-constrained resource allocated to MT $k$  by  $\rho^{(k)}$, e.g., the number of RBs in LTE or channel utilization time in IEEE 802.11p. Further, for MT $k$, we map the tuple (link, MCS, resource allocation), $\{l^{(k)}, \omega^{(k)}, \rho^{(k)} \}, \forall k \in \{1,\cdots, K\}$ in the $n$-th monitoring slot within the same decision-making period into a new, single action. Thus, the action space comprises choices for the selection of the appropriate link, MCS, and the amount of radio resources over the chosen link. We recall  that an action is selected at the end of every decision period, and it is applicable to the subsequent $N$ monitoring slots.  

Next, without loss of generality and for notation simplicity, we focus on two links only, and  denote the number of  different MCS values supported over each link by $i$ and $j$, respectively. Also, we discretize the quantity of radio resources that can be allocated over each link, and indicate them with $p$ and $q$, respectively. Then, the action space is given by $\mathcal{A}:= \{a^{(k)} \in [0,(ip+jq-1)] \}$, such that $a^{(k)} = \{0, \ldots, ip-1 \}$ when the first (e.g., LTE) link is selected, and $a^{(k)} = \{ip, \ldots, ip+jq-1 \}$ in case of the second (e.g., IEEE 802.11p) link. The advantage of such definition of an action is that it limits the action space to a subset of discrete positive integers with low cardinality, and facilitates simultaneous selection of several resources with a single action.

{\textbf {Reward.}} Given a traffic flow, we consider as KPIs the packet loss rate  and the latency observed at the MAC layer during  packet transmission. 
To meet the  KPI requirements at the MT, it is required to provide the traffic flow with radio resources such that the observed KPIs are always less or equal to their target values (hereinafter also referred to as   thresholds). Beside meeting the KPI thresholds, it is essential to keep the observed KPIs as close as possible to the respective KPI thresholds for optimum utilization of network resources: substantially better values than the target ones would indeed translate into a waste of resources. Thus, the choice of reward function should be such that it equally accounts for both the KPIs and its value increases as the observed KPIs approach the corresponding thresholds and vice versa.

Let the observed packet loss rate, target packet loss rate, observed latency, and target latency be denoted with $x_o$, $x_{th}$, $l_o$, and $l_{th}$, respectively. We define the reward value $r$ as the sum of two reward components corresponding to packet loss $r_x(\cdot)$ and latency $r_l(\cdot)$, respectively. 
Thus, for the $k$-th MT, at the $n$-th monitoring slot, we have: 
\begin{equation} \label{eq:rew}
r(s^{(k,n)}, a^{(k)}) = r_x(s^{(k,n)}, a^{(k)})+ r_l(s^{(k,n)}, a^{(k)})
\end{equation} 
where packet loss and latency components are given by:
\begin{subequations}
\begin{align}
r_x(s^{(k,n)}, a^{(k)}) =  1- \text{erf}(x_{th}-x_o) \\
r_l(s^{(k,n)}, a^{(k)}) =  1- \text{erf}(l_{th}-l_o) 
\end{align}  
\end{subequations}  
if the target KPIs are met, and by:
\begin{subequations}
\begin{align}
r_x(s^{(k)}, a^{(k,n)}) =  \text{erf}(x_{th}-x_o) \\
r_l(s^{(k)}, a^{(k,n)}) =  \text{erf}(l_{th}-l_o) 
\end{align}  
\end{subequations}  
otherwise.

\begin{figure*}
\centering
\subfloat[\label{fig:rew_KPI_a}]{\includegraphics[width = 0.24\textwidth]{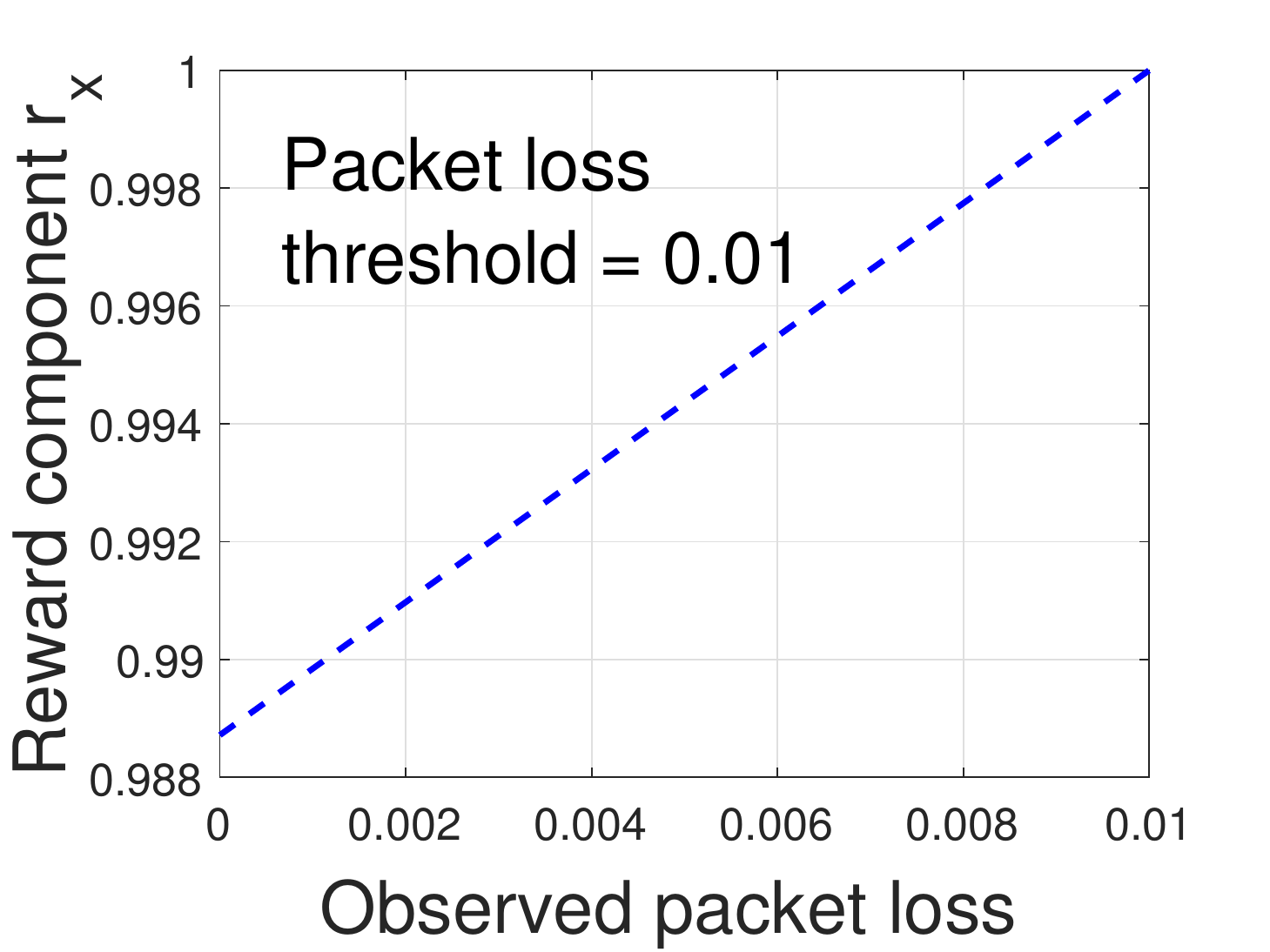}}
\subfloat[\label{fig:rew_KPI_b}]{\includegraphics[width = 0.24\textwidth]{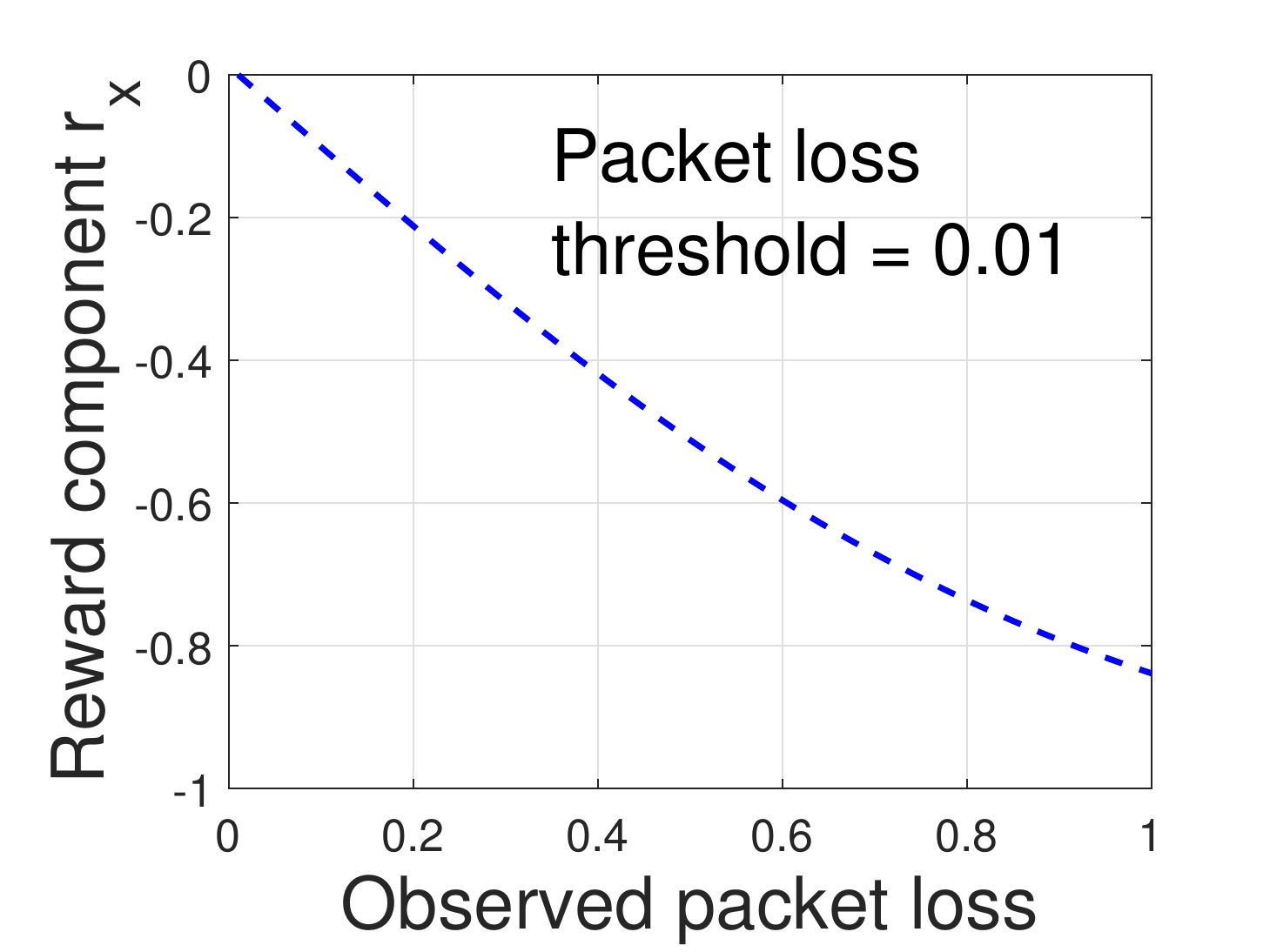}}
\subfloat[\label{fig:rew_KPI_c}]{\includegraphics[width = 0.24\textwidth]{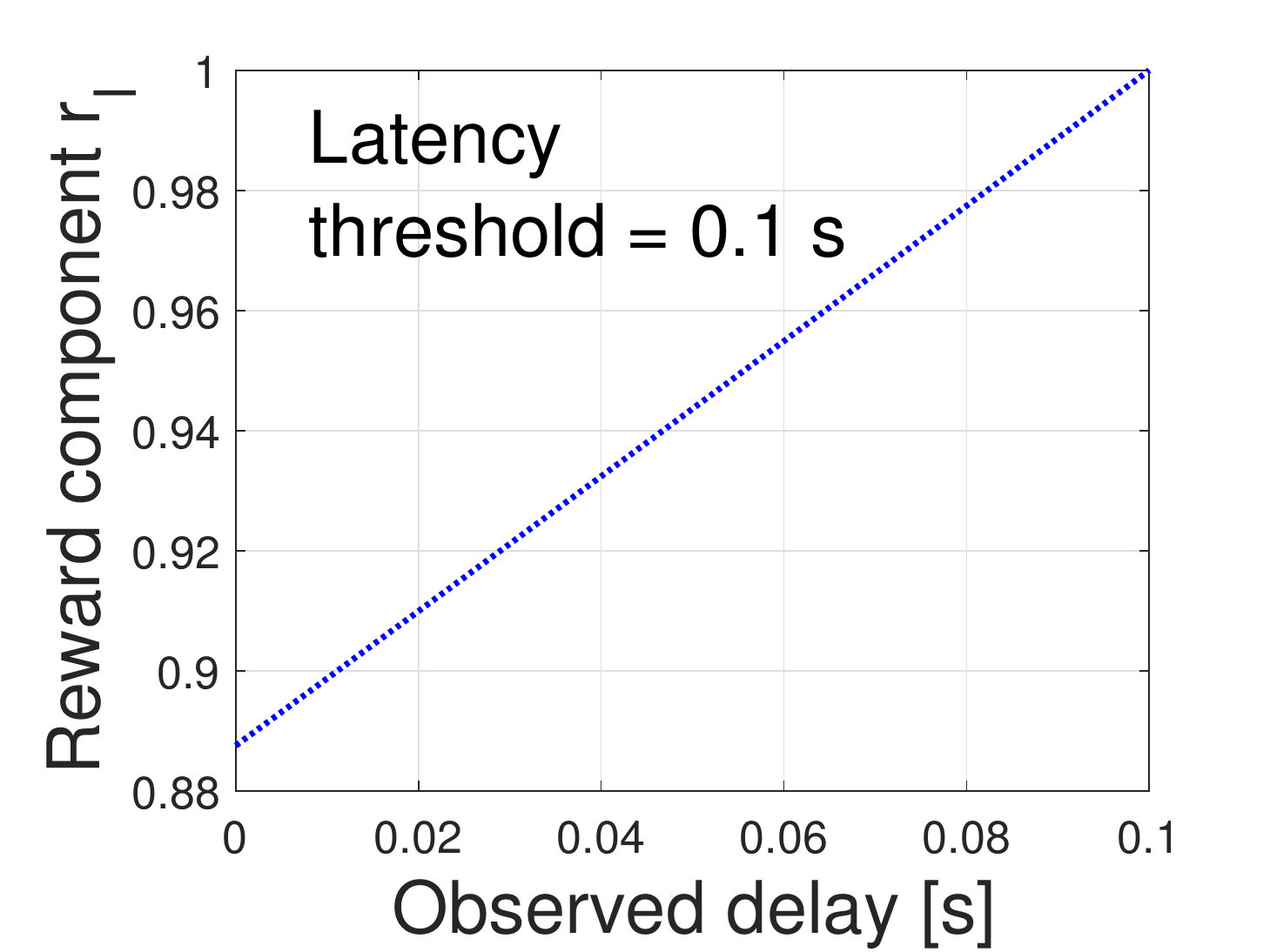}}
\subfloat[\label{fig:rew_KPI_d}]{\includegraphics[width = 0.24\textwidth]{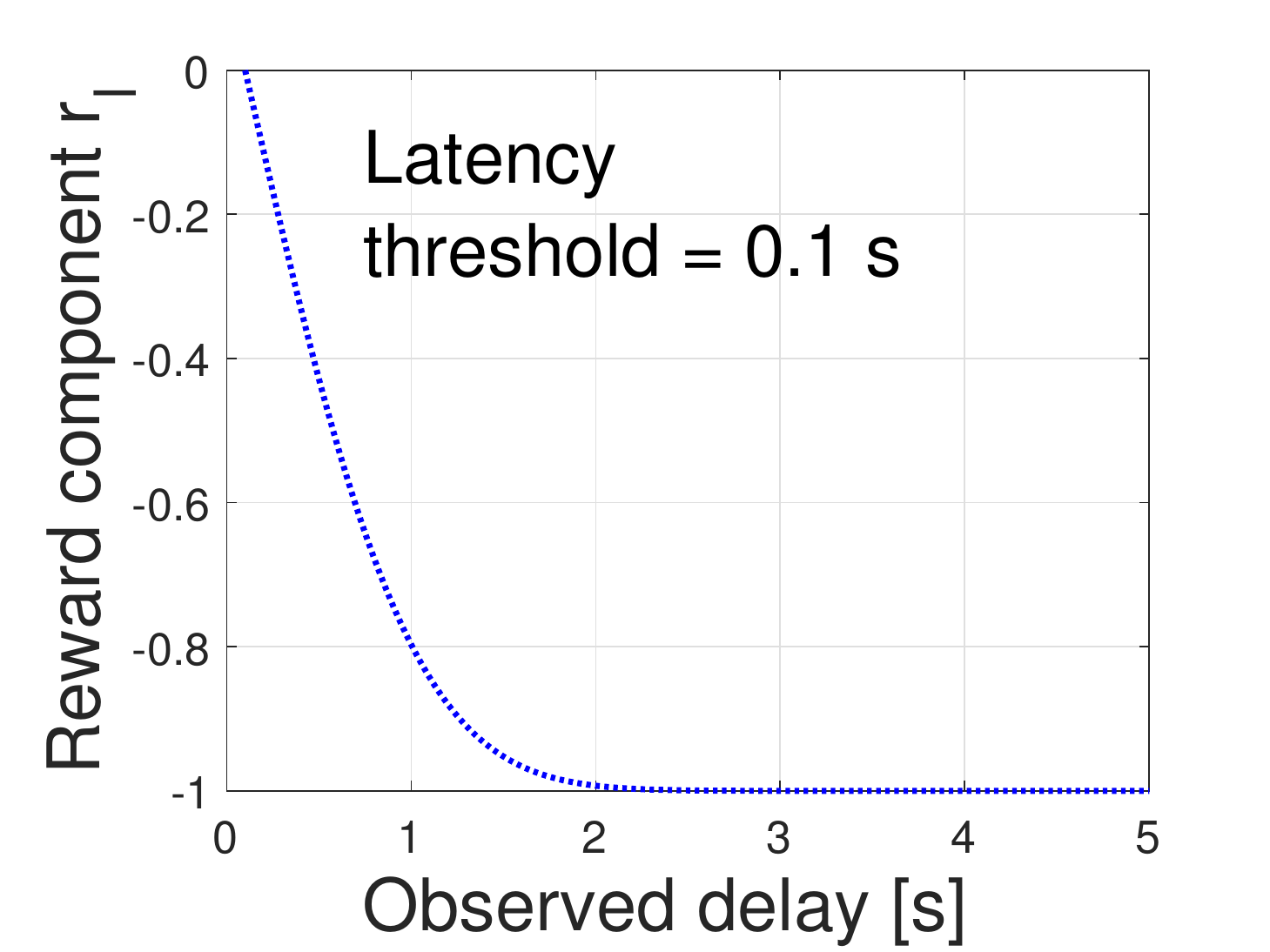}}
\caption{{Variation of reward component as a function of packet loss $r_x$ (fulfilled (a) and  unfulfilled (b) target KPI) and as a function of latency $r_l$ (fulfilled (c) and unfulfilled (d) target KPI). The reward components take positive values as long as the KPI targets are fulfilled, and negative otherwise; further, the closer the KPI to its target value, the higher the reward.}} 
\label{fig:rew_KPI}
\end{figure*}

Since the maximum and minimum value of the $\text{erf}$ function lies between $+1$ and $-1$, we have: $-2 \leq r(s^{(k,n)}, a^{(k)}) \leq 2 $. Our choice of $\text{erf}$ for estimating individual reward components is motivated by its shape, which takes $0$ value at the origin, and gradually increases (decreases) and saturates to the maximum (minimum) value in the positive (negative) direction. Consequently, for the individual reward components, in the positive region of operation, i.e., when the KPI threshold is met, the reward value is positive and it further increases to its maximum value at $+1$ as the observed KPI approaches its target KPI value. Likewise, in the negative region of operation, i.e., when the KPI threshold is not met, the value of the individual reward components is negative, which further reduces and saturates to the minimum value $-1$ as the observed KPI moves away from the KPI threshold. 
Note that the penalty in the reward component for observed KPI overshooting the KPI threshold is larger than for observed KPI undershooting the KPI threshold by the same amount. This is so because an observed KPI value that exceeds the KPI threshold adversely affects the QoS at the user end, which is a critical issue from the system design point of view. On the contrary, the case of an observed KPI undershooting the KPI threshold is an acceptable situation, however even in this case, we do not want the observed KPI value to deviate from the KPI threshold by a large amount, as this would lead to the system performing exceptionally well at the cost of extra resource consumption.  
The variation of the reward components as a function of the observed KPI values for packet loss and latency is shown in Fig.\,\ref{fig:rew_KPI}, which highlights how the behavior of the individual reward components in the typical range of observed KPI values is as desired.

We recall that the goal of the RL agent is to eventually maximize the cumulative reward measured as the sum of immediate reward and future rewards in the long run. 
To this end, we consider the  generic decision period $h$ and, extending the previous notation, we let $a^{(k,h-1)}$ denote the action  for  MT $k$ selected in decision period $(h-1)$ and applied in decision period $h$. We then 
define the average reward over $h$ as,
{
\begin{equation} \label{eq:avg_rew}
\overline{r}(\sv^{(k,h)}, a^{(k,h-1)} ) = \frac{\sum_{n=1}^{N} r(s^{(k,n)}, a^{(k,h-1)})}{N}
\end{equation} }
where 
$\sv^{(k,h)}$ is the vector of states observed for MT $k$ in the $N$ monitoring slots in decision period $h$, while   $a^{(k,h-1)}$ is the  action  for  MT $k$ selected in decision period $h-1$ and applied in decision period $h$. 
Then, we adopt the definition of cumulative reward for the $k$-th MT, observed during  decision period $h$, as the differential return $G^{(k,h)}$  \cite{sutton},
{
\begin{equation}
G^{(k,h)} = \sum_{\ell=0}^\infty \overline{r}(\sv^{(k,h+\ell)}, a^{(k,h+\ell-1)})-(l+1)\widehat{r}(k,\pi) 
\end{equation} }
where $\pi:\mathcal{X} \rightarrow \mathcal{A}$ denotes the radio policy mapping the context space of each MT into actions, and $\widehat{r}(k,\pi)$ in  \cite{sutton}:
\begin{equation}\label{eq:r-k-pi}
\widehat{r}(k,\pi)= \lim_{h\to\infty} \frac{1}{h} \sum_{t=1}^h \mathbb{E}[
\overline{r}(\sv^{(k,t)}, a^{(k,t-1)}) | \overline{s}^{(k,1)}, a^{(k,0)} \sim  \pi] \,.
\end{equation}
In (\ref{eq:r-k-pi}), we consider $h=0$ to be the time at which the algorithm execution started, and  $\overline{s}^{(k,1)}$ is the mean state computed averaging over the state values observed in the $N$ monitoring slots of  the initial decision period. 
Thus, 
$\widehat{r}(k,\pi)$ is obtained as the average of the reward conditioned on $\overline{s}^{(k,0)}$ and the 
subsequent actions taken according to policy $\pi$.

{\textbf {Action value estimation.}} Given decision period $h$, at the end of the corresponding $N$ monitoring slots,  actions need to be evaluated in order to ultimately select the best one. 
To this end, we compute the average context  over the $N$ monitoring slots in $h$ for each given MT $k$, as
{
\begin{equation} \label{eq:avg_state}
\overline{s}^{(k,h)} = \sum_{n=1}^{N} \frac{y_n s^{(k,n)}}{\sum_{n=1}^{N} y_n}
\end{equation} }
where $y_n > 0$ and $y_N > y_{N-1} > \cdots y_1$  are the weights assigned such that the latest context has the highest weight. Although they can be arbitrarily set, in our experiments we fix them to $1, \ldots, N$, in accordance with the temporal sequence of the monitoring slots.
We then quantify
the goodness of taking an action in such a context using action values.  
For MT $k$, if $a^{(k,h)}$ is selected based on state $\overline{s}^{(k,h)}$ under policy $\pi$, then its action value $q_{\pi}(\overline{s}^{(k,h)},a^{(k,h)})$ is defined as expected differential return conditioned on  $\overline{s}^{(k,h)}$ and $a^{(k, h)}$, following policy $\pi$. Mathematically,
\begin{equation}
q_{\pi}(s,a) = \mathbb{E}_{\pi} [ G^{(k,h)} | \overline{s}^{(k,h)} = s, a^{(k,h)} = a] \,.
\end{equation}
Apparently, a policy, $\pi$, can be better than any other policy $\pi'$ if $q_{\pi}(s,a) \geq q_{\pi'}(s,a) $.  Since the context vector comprises SNR, link aggregate traffic load and buffer state, context space $\mathcal{X}$ is real and an uncountable number of states are possible. Consequently, tracking action values corresponding to different contexts is not scalable. To overcome this problem, we use a practical method for action value estimation using function approximation in an F-dimensional space, {yielding  the following approximated  function,
\begin{equation} \label{eq:tile_coding_approx}
\hat{q}_{\pi}(\overline{s}^{(k,h)},a^{(k,h)},w)=\sum_{f=1}^F w_f x_f(\overline{s}^{(k,h)}, a^{(k,h)})
\end{equation}} 
where $\wv=[w_1,\dots,w_F] \in \mathbb{R}^F$ and $x_f(\overline{s}^{(k,h)}, a^{(k,h)})$ denote the weight and feature vector, respectively. Here, feature vector $x_f(\overline{s}^{(k,h)}, a^{(k,h)})$ 
is generated using tile coding \cite{tile_coding}, which converts a point in the 2-dimensional context vector into a binary feature vector such that vectors of neighboring points have a high number of common elements. 
The continuous space of context variables is tucked up with tiles,  and each tile corresponds to an index in the binary feature vector. Several offset grid of tiles, called tilings, are then stacked over the space to create regions of overlapping tiles. We have used $8$ tilings, with $512$ tiles each. Every context vector falls  in one tile in each of the $8$ tilings, which correspond to $8$ features.

{\textbf {Action selection.}} The estimation of the action values is followed by an $\epsilon$-greedy action selection policy \cite{sutton}, which selects the best action for each MT so as to maximize its cumulative reward over an infinite time horizon. We consider an $\epsilon$-greedy action selection with $\epsilon = 0.5$ and $\epsilon$-decay factor $= 0.999$. Thus, for MT $k$, if the average context over  decision-making period $h$, $\overline{s}^{(k,h)}$, and the action value estimates for all possible actions 
$a^{(k,h)} \in \mathcal{A}$ in $\overline{s}^{(k,h)}$ are obtained as $\hat{q}_{\pi}(\overline{s}^{(k,h)},a^{(k,h)},\wv)$, the greedy action for the MT, $\tilde{a}^{(k,h)}$, is chosen with probability $1- \epsilon$ such that $\tilde{a}^{(k,h)} = \text{argmax}_a {\hspace{+0.5mm}}\hat{q}_{\pi}(\overline{s}^{(k,h)},a^{(k,h)},\wv)$. 
The $\epsilon$ parameter decays by a factor of 0.999 in the subsequent decision period. This favors higher exploration while the environment is still unfamiliar; with progression of time, instead, it allows for further exploitation of the environment knowledge gained during the exploration, so as to maximize the expected return.

\subsection{Pareto block} 
At the input of the CAREM framework, contexts from different MTs are considered independently from each other, which may sometimes lead the radio policy to choose actions for different MTs where the sum of individual allocations of a capacity-constrained resource exceeds its respective maximum availability. We solve this issue by introducing a novel algorithm that further fine tunes the resources allocated by the radio policy. Such refinement  makes sure  that if multiple MTs are using the same link, the allocation is Pareto-efficient fair and the sum of allocated resources to the MTs adheres to the maximum capacity constraint of the link. 

We focus again on a given decision period, thus omitting the dependency on $h$, and we map the Pareto-efficient fair allocation of a capacity-constrained resource on a given link $l$ across all MTs assigned to $l$, onto a multi-criteria optimization problem as detailed below. 
Let us denote the set of MTs assigned to link $l$ by ${\Kc_l}$. Given a set of coefficients $v_k \geq 0, k \in \Kc_l$, such that $\sum_{k\in \Kc_l} v_k = 1$, it is required to find a solution  $S=\{\rho^{(k)}\}_{k\in \Kc_l}, S \in \Phi$, that maximizes $\sum_{k \in \Kc_l} v_k\Gamma^{(k,n)}(S)$ such that $\sum_{k \in \Kc_l} \rho^{(k)} \leq \rho_{max}$. Here, $\Phi$ is the set of feasible solutions, $\rho^{(k)}$ is the capacity-constrained resource allocated to the $k$-th MT during the considered decision-making period, $\Gamma^{(k,n)}(S) $ is the criteria function denoting the reward of  MT $k$ in the $n$-th monitoring slot following the resource allocation strategy $S$, and $\rho_{max}$ is the maximum availability of the capacity-constrained resource. 

{The optimization problem is solved using an iterative multi-objective search and update algorithm described as follows. The Pareto block is invoked if the sum of allocated capacity-constrained resource across all MTs exceeds $\rho_{max}$. To start with, the capacity-constrained resource $\rho^{(k)}$ is extracted from the greedy action $\tilde{a}^{(k)}$ to form solution $S_1 = \{\rho^{(k)}\}_{k\in \Kc_l}$. Then  each $\rho^{(k)}$ in $S_1$ is scaled so that $\sum_{k \in \Kc_l} \rho^{(k)} (S_1) \leq \rho_{max}$. Note that, beside $S_1$, other solutions $S_i$ are possible as well where $\rho^{(k)}(S_i) \geq \rho^{(k)}(S_1)$, as intuitively, if an action $\tilde{a}^{(k)} = \{l^{(k)}, \omega^{(k)}, \rho^{(k)} \}$ is feasible for a given $\rho^{(k)}$, it will also be feasible for any other allocation $\rho'^{(k)} \geq \rho^{(k)}$.  In such case, $ \sum_{k\in \Kc_l} \rho^{(k)}(S_i) \geq  \rho_{max}$, however since scaling is anyways imperative to satisfy the maximum capacity constraint, it is in best interest to consider all such possible solutions. In view of this argument, we create an expanded solution set ${\cal S}_e = \{ S_1, S_2, \ldots \}$ such that $\forall S_i \neq S_1$, 
\begin{equation} \label{eq:exp_sol_set}
\rho^{(k)}(S_i) \geq \rho^{(k)}(S_1), \forall k \in \Kc_l, \wedge  \sum_{k\in \Kc_l} \rho^{(k)}(S_i) \leq |\Kc_l| \rho_{max}\,.
\end{equation}
Further, we create scaled expanded solution set ${\cal S}_s$, 
\begin{equation} \label{eq:scal_exp_sol_set}
{\cal S}_s = \{S_i/|{\Kc_l}|\}_{{\cal S}_e}\,.
\end{equation}
Subsequently, Pareto dominant solution set $\mathcal {S}$ is obtained through an iterative search and update over ${\cal S}_s$ using the following condition $\forall S' \in \mathcal{S}$ and $S'' \in \Sc_s$, 
\begin{equation} \label{eq:pareto}
\Gamma^{(i)}(S') > \Gamma^{(i)}(S''), \Gamma^{(j)}(S') \geq 	\Gamma^{(j)}(S''), \forall i,j \in {\Kc_l}, i \neq j\,.
\end{equation}
Finally, an optimal solution $S^* = \{\rho^{*(k)}\}_{k\in \Kc_l} $ is chosen from the Pareto dominant solution set $\mathcal{S}$ such that $\forall S' \in \mathcal{S}, S^* \in \mathcal{S}$,
\begin{equation} \label{eq:fair_sol}
\max \underset{i \in \Kc_l}{\min}  (v_i \Gamma^{(i)}(S^*)) \geq \max
\underset{i \in \Kc_l}{\min}(v_i \Gamma^{(i)}(S')) \,.
\end{equation}
Since $S^*$ maximises the minimum criterion function across all $k\in \Kc_l$, it is the required Pareto-efficient fair solution. For each link, the procedure to obtain such solution, i.e., action $a^{*(k)}$ for each MT $k$, is summarized in Algorithm \ref{algo:Pareto}. Using theorems defined in \cite{laumanns2002}, it can be proved that the obtained solution is Pareto efficient.}

\floatname{algorithm}{Algorithm}
\begin{algorithm}[tb]
\caption{\label{algo:Pareto} 
Allocation of the capacity-constrained resource on radio link $l$ 
}
\begin{algorithmic}[1]
  \State{Input: greedy actions,  $\tilde{a}^{(k)} = \{l^{(k)}, \omega^{(k)}, \rho^{(k)} \}, \forall k\in \Kc_l$ }
\State{Extract capacity-constrained resource allocation $S_1 = \{\rho^{(k)}\}_{k\in \Kc_l}$ from greedy actions} 
{ \If{$\sum_{k \in \Kc_l} \rho^{(k)} \leq \rho_{max}$}
 \Comment{Check on the resource allocations on link $l$}
\State{$S^*  \gets S_1$}
        \Comment{Pareto-efficient fair solution}
        \Else
        \State{Create expanded solution set, ${\cal S}_e$ using (\ref{eq:exp_sol_set})  }
        \State{Rescale ${\cal S}_e$ to create ${\cal S}_s$ using (\ref{eq:scal_exp_sol_set})}
        \State{Identify Pareto dominant solution set $\mathcal {S}$ using (\ref{eq:pareto})}
        \State{Compute Pareto-efficient fair solution $S^*$ using (\ref{eq:fair_sol})}
        \EndIf }
	\State{Output: Pareto-efficient fair actions,  $a^{*(k)} = \{l^{(k)}, \omega^{(k)}, \rho^{*(k)} \}, \forall k \in \Kc_l $ } 
\end{algorithmic}
\end{algorithm}
\vspace{-3mm}

\subsection{Learning algorithm}
In the absence of any prior knowledge of the environment, here we exploit the concept of experience-based learning using sample sequences of context, actions, and rewards observed from the actual interaction of the RL agent with the environment. SARSA, an acronym for quintuple ($S_t, A_t,R_t, S_{t+1}, A_{t+1}$), is an on-policy algorithm where learning of the RL agent at time $t$ is governed by its current state $S_t$, choice of action $A_t$, reward $R_t$ received on taking action $A_t$, state $S_{t+1}$ that the RL agent enters after taking action $A_t$, and finally the next action $A_{t+1}$ that the agent chooses in new state $S_{t+1}$ \cite{sutton}. 
Then, 
given the average context vectors for the different MTs  and the possible actions, the key steps involved in the learning of the SARSA approach are: (i) estimation of action values $q_{\pi}(s,a)$, (ii) action selection for each MT, (iii) Pareto-efficient fair action tuning,   and (iv) update of the action-value estimates.

{\textbf {Action value update.}} Action values satisfy the recursive Bellman equations given as,
\begin{equation}\label{eq:q_pi}
q_{\pi}(s,a)=\sum_{r,s'}p(s',r|s,a)[r-\widehat{r}(k,\pi)+\sum_{a'}\pi(a'|s')q_{\pi}(s',a')] 
\end{equation}
where 
{
\begin{eqnarray}
p(s',r|s,a)&\hspace{-0.7cm}\mathord{=}\hspace{-0.7cm} &\PP\Big\{\overline{s}^{(k,h)} = s', \overline{r}(\sv^{(k,h)},a^{(k,h-1)}) =  r| \Big . \nonumber\\
&&\Big . \hspace{0.5cm}\overline{s}^{(k,h-1)}=s,a^{(k,h-2)}= a \Big\}\,,
\end{eqnarray} } 
with $\pi(a'|s')$ being the probability of taking action $a'$ in state $s'$ under policy $\pi$. This fundamental property forms the basis of the update of the action values of the present context, based on an error term defined as the difference between a target action value and the current action value. Details on Bellman equation and the derivation of the update rule can be found in \cite{sutton}.   Here we consider the temporal difference learning, 
in which the target action value for the context in the given  decision period $h$ is the bootstrapping estimate of action values for the context in the subsequent  decision period $(h+1)$. 
Since the difference in action value estimates of successive contexts drives the learning procedure, error is termed as temporal difference error $\delta$, given by

{\begin{equation}  \label{eq:delta}
\begin{aligned}
 \delta = \overline{r}(\sv^{(k,h)}, a^{*(k,h-1)}) -\widehat{r}(k,\pi) + \hat{q}_{\pi}(\overline{s}^{(k,h+1)},\\ a^{*(k,h+1)}, \wv) 
 -\hat{q}_{\pi}(\overline{s}^{(k,h)}, a^{*(k,h)},\wv)\,.  
 \end{aligned}
 \end{equation}}
In (\ref{eq:delta}), 
 $a^{*(k,h)}$ is the Pareto-efficient fair action  for  MT $k$ selected in decision period $h$ and applied in decision period $(h+1)$. 
 Also, we recall that  $\overline{s}^{(k,h)}$ and $\overline{r}(\overline{s}^{(k,h)}, a^{*(k,h-1)})$ are (resp.) the weighted mean context and mean reward observed over decision period $h$.
 
Subsequently, $\delta$ is used to update $\widehat{r}(k,\pi)$ and weight vector $\wv$ using gradient descent as, 
\begin{equation} \label{eq:update_rew}
\widehat{r}(k,\pi) \leftarrow \widehat{r}(k,\pi)+\beta\delta
\end{equation} 
\begin{equation} \label{eq:update_w}
\wv \leftarrow \wv + \alpha\delta \nabla \hat{q}(\overline{s}^{(k,h)}, a^{*(k,h)},\wv) 
\end{equation}
where $\alpha$ and $\beta$ are the step sizes for updating weight vector and average reward conditioned on initial state and the subsequent actions, respectively.  Note, however, that the bootstrapping target itself depends on the weight vector. Consequently, it is biased and does not produce a true gradient descent, hence this is referred to as a semi-gradient method. Step sizes $\alpha$ and $\beta$ govern the learning rate of the algorithm by deciding how much closer the estimate moves towards the target in a single iteration. If the chosen step sizes are too small, it takes a large time to reach the best values of weights and average reward, hence slower learning, which is clearly undesirable. On the contrary, although a large step size may reduce the training time, there is a possibility of overshooting the true optimum position and oscillating between local optima. To this end, in our experiments, we  considered different choices and found $\alpha, \beta = 0.01$ to be the best suited one.

The workflow of the \CaRReM\ RL algorithm is summarized in Algorithm\,\ref{algo:RL}, where for simplicity we focus on MT $k$ and decision period $(h+1)$. Parameters including the decision-making periodicity, $N$, and step sizes, $\alpha$ and $\beta$, are initialized at the start of the algorithm. Given a decision period $(h+1)$, after observing the context vector, the radio policy gives as  output a greedy action for MT $k$. 
Once Algorithm\,\ref{algo:RL} is run for all MTs, the greedy actions are further tuned by the Pareto block to obtain Pareto-efficient fair actions $\{a^{*(k,h+1)}\}$, subsequently reinforcement learning takes place using differential semi-gradient SARSA.
Specifically, the temporal difference error  $\delta$, the average reward conditioned on initial state and subsequent actions, and the weight vector are updated using 
(\ref{eq:delta}), (\ref{eq:update_rew}), and  (\ref{eq:update_w}), respectively.  
Note that, although the SNR and buffer state for MT $k$ may be independent of those experienced by other MTs, due to the sequential decision making process, information of the link allocated to  MT $k$ is used to update the aggregate link load on each link while making decisions for subsequent MTs.
\floatname{algorithm}{Algorithm}
\begin{algorithm}[tbh]
\caption{\label{algo:RL}RL algorithm in \CaRReM\ for MT $k$ in decision period $(h+1)$
}
\begin{algorithmic}[1]
 \State{Define parameters: decision-making periodicity  $N$, step sizes $\alpha, \beta \in (0,1]$} 
 	\For{the $n$-th monitoring slot in the $h+1$-th decision period, $n = 1,2, \cdots, N $}
 	\If{$n=1$}
        \State{
        $s^{(k,n)} \gets \overline{s}^{(k,h)}, a^{*(k)} \gets a^{*(k,h)}$}
        \Else
        \State{Observe $s^{(k,n)}, a^{*(k)}\gets a^{*(k,h)} $}
        \EndIf
        \State{{Evaluate reward per slot $r(s^{(k,n)}, a^{*(k)} )$ using (\ref{eq:rew})}}
  \EndFor	
  \State{{Find mean reward over the $h+1$-th decision period, $\overline{r}(\sv^{(k,h+1)}, a^{*(k,h)} )$  using (\ref{eq:avg_rew}) }   
  \State{Find weighted mean context  $\overline{s}^{(k,h+1)}$ using (\ref{eq:avg_state})}
  \State{Compute action values $\hat{q}_{\pi}(\overline{s}^{(k,h+1)}, \cdot,\wv)$ for all possible actions using (\ref{eq:tile_coding_approx})}}
\State{Choose $\tilde{a}^{(k,h+1)}$  using the $\epsilon$-greedy policy}
\end{algorithmic}
\end{algorithm}
\vspace{-3mm}

\subsection{Computational complexity analysis}
Based on Algorithms\,1 and 2, the most complex operations are given by the following steps: (i) greedy action selection for the $K$ MTs, (ii) Pareto-efficient fair action selection on $L$ links, (iii) computation of weighted mean of context and mean reward per decision period for the $K$ MTs, and (iv) update of weight vector for learning radio policy based on KPI observation from the $K$ MTs. Corresponding to each of these steps and considering that the number of radio resources is finite, the computational complexities are given by  $\mathcal{O}(K |\mathcal{A}|)$, 
$\mathcal{O}(K)$,
$\mathcal{O}(K)$, and $\mathcal{O}(K N)$, respectively. 
Hence,  the overall complexity is given by $\mathcal{O}(K |\mathcal{A}|) + 
\mathcal{O}(K) 
+ \mathcal{O}(K)+ \mathcal{O}(K N)
\approx \mathcal{O}(K|\mathcal{A}|) + 
 \mathcal{O}(K N) $, where the first term is the dominant one as  $|\mathcal{A}|$ is much larger than $N$.

We recall that every decision period is comprised of $N$ monitoring slots. The smaller the value of $N$, the more frequently CAREM makes decisions and the more the performed computations, with worst case scenario given at $N=1$ wherein computations on the order of $\mathcal{O}(K |\mathcal{A}|) + \mathcal{O}(K) $  are done every monitoring slot. It follows that choosing a high $N$ value  leads to a significant computation gain,  at the cost (as shown in Sec.\,\ref{results}) of a marginal performance degradation.

\begin{figure*}[tb]
\center
\includegraphics[width=0.7\textwidth]{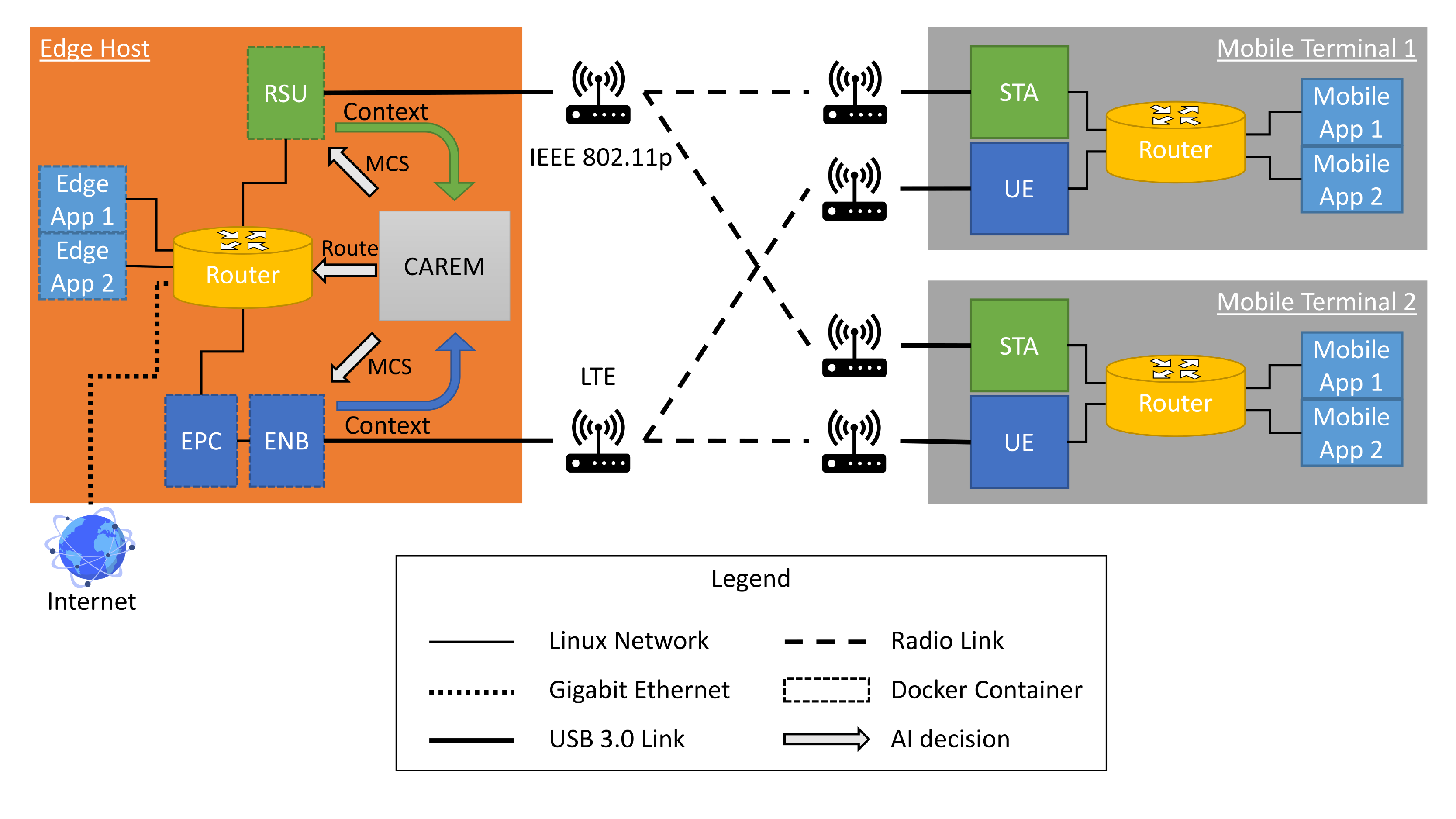}
	\caption{Testbed architecture (for clarity two links and two mobile terminals only are shown).
	{ Edge Host provides  connectivity to the MTs through a heterogeneous vRAN. For clarity, two links (3GPP LTE and IEEE 802.11p) and two MTs only are shown.}}
	\label{fig:testbed_arch}
	\vspace{-3mm}
\end{figure*}

\section{Testbed Design and Implementation} \label{testbed_design}

The testbed architecture, illustrated in Fig.\,\ref{fig:testbed_arch}, is composed of two main interconnected blocks: the edge host (left block) and the mobile terminal (right block). For clarity of presentation, only two links and two MTs are shown. The purpose of the edge host is to provide computational resources and mobile connectivity for services offered by the edge applications, which are then consumed by the mobile applications running at the MTs. Connectivity between the edge host and the MTs is provided through a heterogeneous vRAN integrating the 3GPP LTE (bottom link in Fig.\,\ref{fig:testbed_arch}) and IEEE 802.11p (top link) technologies, both implemented through SDR solutions.

The LTE vRAN is based on srsRAN \cite{miguelez}, an open-source SDR LTE stack implementation that offers EPC, eNB, and MT applications. It is compliant with LTE Release 9 and supports up to 20\,MHz bandwidth channels as well as transmission modes from 1 to 4, all using the FDD configuration.
The IEEE 802.11p transceiver 
is implemented through a GNU Radio flowgraph, released by the WiME project \cite{bloessl}, and it is interoperable with commercial IEEE 802.11p devices.  We mention here that, because of processing delay limitations, the IEEE 802.11p transceiver lacks important features of the standard such as ACKs and CSMA/CA mechanisms.

\begin{figure}[!t]
\centering{{\includegraphics[width=0.9\linewidth]{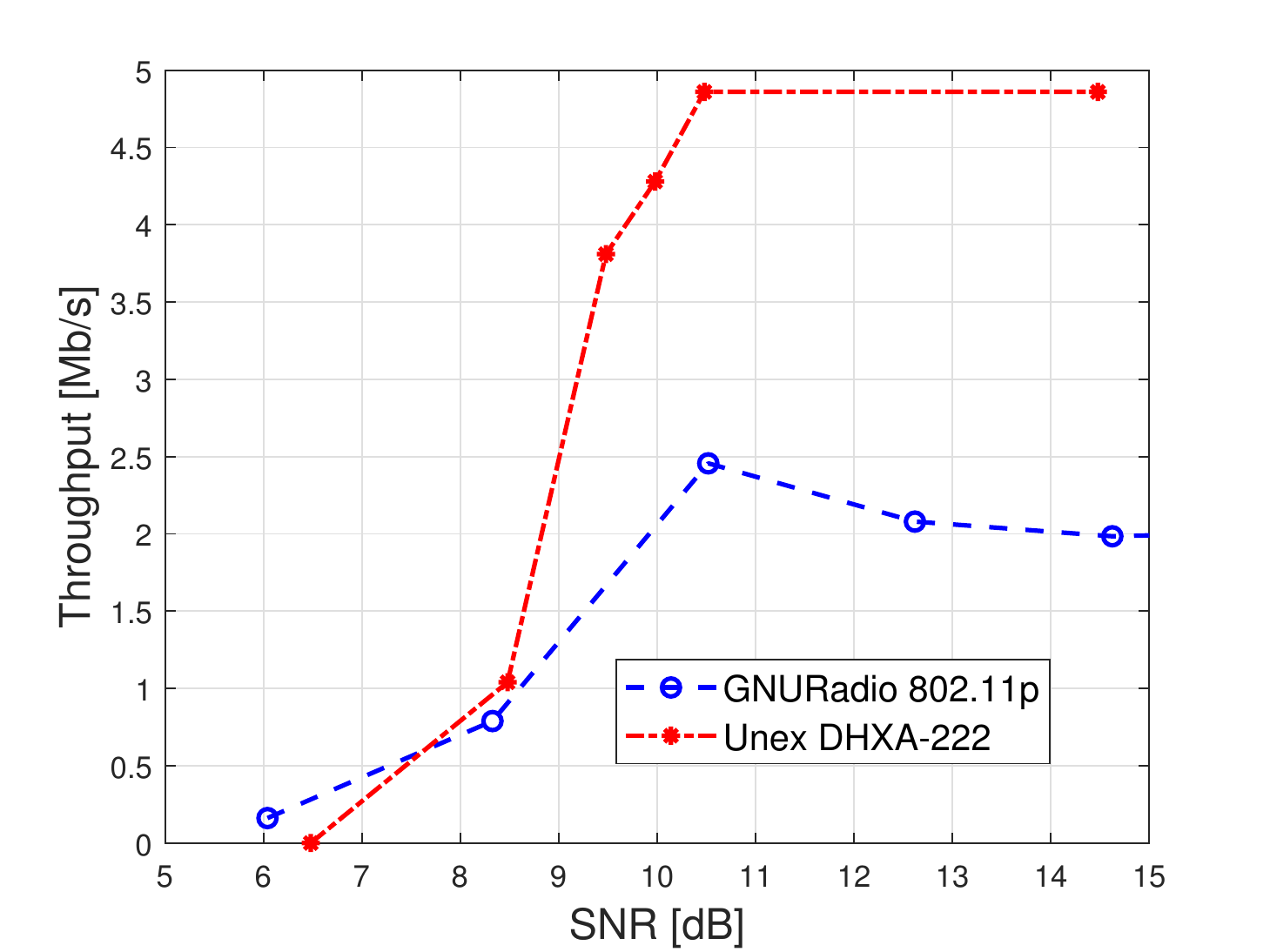} }}
   \caption{Throughput of IEEE 802.11p with MCS set to 2:  comparison  between the SDR interface implemented through the GNURadio card (blue curve) and that of the off-the-shelf device, namely, Unex card (red curve).
   {The SDR IEEE 802.11p transceiver performance has been properly scaled to match that of commercial cards.}}
\label{fig:11p_throughput_comparison}
\vspace{-3mm}
  \end{figure}

The core component of edge host is the proposed \CaRReM\ framework, which controls the operation of the heterogeneous vRAN. The algorithm periodically selects the appropriate link, MCS to be used, and allocated resources on the selected link for downlink packet transmission. To interact with the host operating system network stack, both the SDR solutions expose a tun/tap interface to which an IP address is assigned. A router is connected to those interfaces to steer traffic over the radio links, the host applications, and the internet, according to the link selected by \CaRReM. The link selection is enforced with dynamic modification to the Linux kernel routing table.

The SDR applications, as well as the edge applications, are implemented and executed within docker containers, to control resource usage and isolate the different applications. The SDR applications have been patched to allow for the dynamic selection of the MCS used for the data radio transmission, according to the radio policy. The srsRAN eNB application has been patched to run a dedicated thread that listens to and applies the MCS and the RBs allocation selected by  \CaRReM\  to the  communication with a specific MT. As for IEEE 802.11p, the GNU Radio flowgraph has been modified  by adding an XMLRPC server block, which exposes a remote procedure call interface to dynamically set the MCS to be used. The airtime allocation instead has been implemented by limiting the IP flow throughput  with Linux Traffic Control according to the radio policy. 
Indeed, since the transceiver does not support the CSMA/CA mechanism, the physical throughput is known given the MCS. 
Furthermore, both the SDR applications have been modified in order to collect such context data as the average SNR and the buffer state report through a sidelink connection. 

The UDP throughput of the SDR IEEE 802.11p transceiver has been compared to the throughput of a commercial wireless card, namely, the Unex DHXA-222, based on the Qualcomm Atheros AR9462 chipset. Using the same MCS (MCS 2: QPSK, 1/2), different levels of SNR have been tested. As shown in Fig.\,\ref{fig:11p_throughput_comparison}, 
the two solutions exhibit similar throughput for SNR below 8.5\,dB, where a high packet loss is observed. At higher SNR, i.e., above 10\,dB, the maximum achievable throughput is instead limited by the physical data rate. Throughput saturates at around 2\,Mb/s using the SDR transceiver -- a value that is more than 50\% lower than the Unex card throughput. Consequently, for the sake of fair comparison between the IEEE 802.11p and the LTE technology, the measured packet loss of the SDR IEEE 802.11p transceiver has been  scaled so that its throughput (and packet loss) performance  matches that exhibited by  commercial cards.

\begin{figure}[!t]
\centering
	\includegraphics[width=0.8\linewidth]{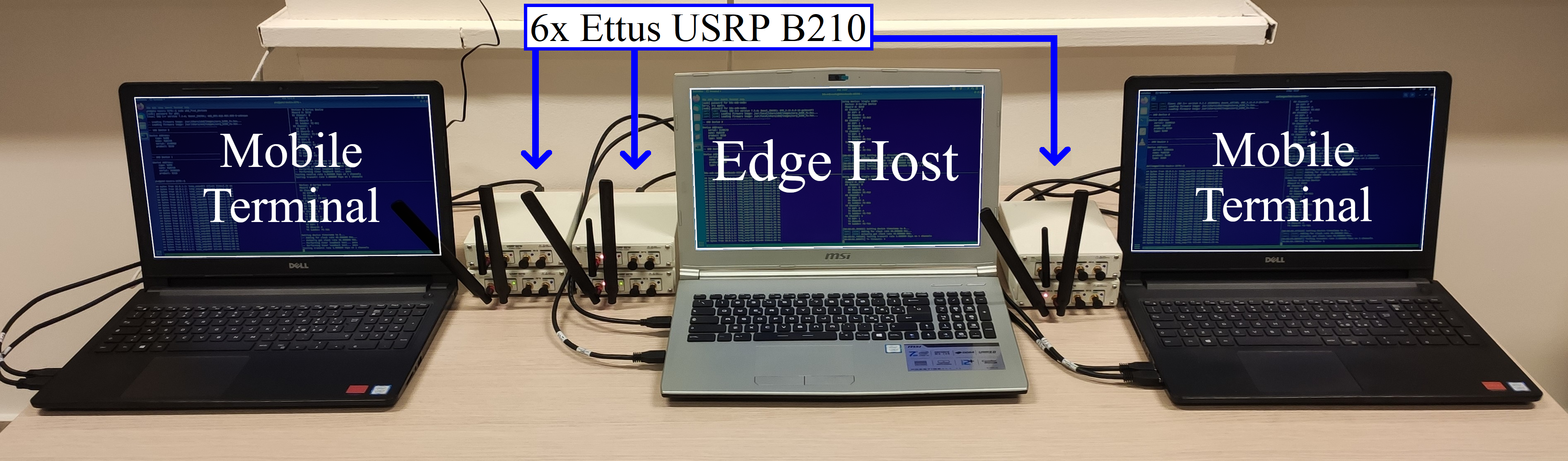}
	\caption{Testbed implementation setup {with two MTs and one Edge Host: each MT is connected to two USRP B210 boards, implementing LTE and  IEEE 802.11p, respectively.}}
	\label{fig:testbed_image}
	\vspace{-3mm}
\end{figure}

The testbed implementation setup used to evaluate  \CaRReM\ is shown in Fig.\,\ref{fig:testbed_image}. The edge host and the MTs are installed in  Ubuntu 18.04 systems. The edge host system is equipped with an Intel i7-7700HQ 4-core CPU and 16\,GB of DDR4 RAM, while the one used for the MTs integrates an Intel i7-8550U 4-core CPU and 16\,GB of DDR4 RAM. Each Ubuntu system is connected to two ETTUS Universal Software Radio Peripheral (USRP) B210 boards, one for LTE and the other for IEEE 802.11p, using USRP Hardware Driver (UHD) v3.15.

\section{Performance Evaluation} \label{results}

In this section, we first detail the experimental settings of the testbed under which we derived our performance results. We then assess the performance of CAREM by showing the convergence of reward values and the behavior of the KPIs in response to the action selection. Finally, we present a comparison of \CaRReM\ with the closest competitive technique  in \cite{romero}, a relatively simpler contextual Bandit (CB) approach, and the standard LTE  cellular system.

\begin{figure}
\centering
\subfloat[]{
  \includegraphics[width=0.8\linewidth]{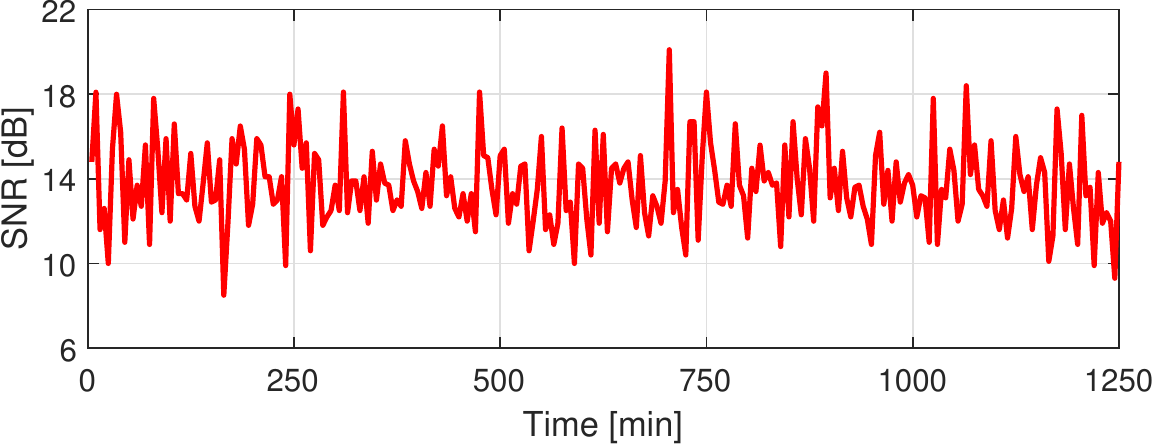}}\\
\subfloat[]{\includegraphics[width=0.8\linewidth]{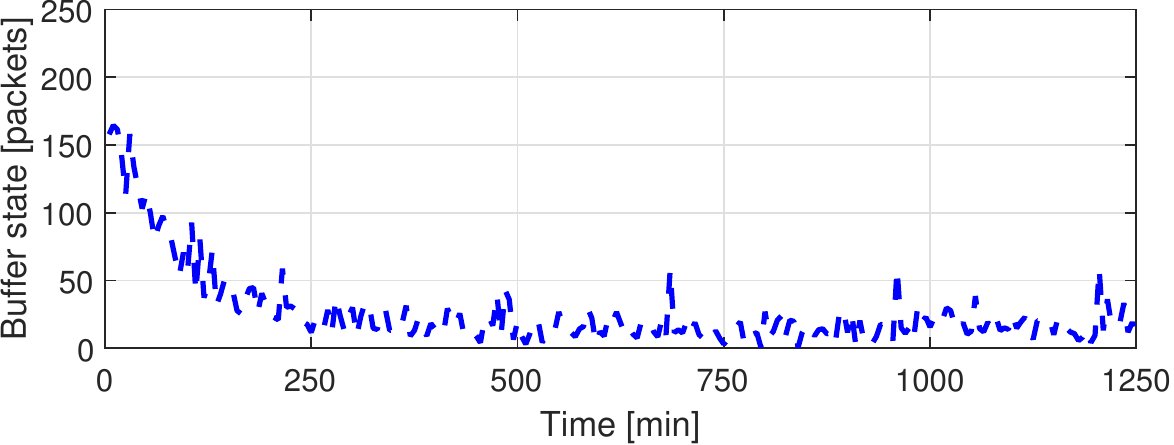}}
  \caption{Variation of context with time: SNR (a) and per-MT buffer state for 3\,Mbps traffic load (b). SNR being an independent variable varies randomly, while  the  buffer state reduces close to zero as the algorithm learns to select better actions.}
\label{fig:context_vs_time}
  \vspace{-3mm}
\end{figure}

\subsection{Experimental settings}
We evaluate the performance of the \CaRReM\ framework using our testbed implementation.  We consider two cases: (a) $N = 1$, which corresponds to per-slot (i.e., 100-ms) decision making, and $N = 10$, where decision is periodically made every $10$ monitoring slots (i.e., every second). 

In our performance evaluation, we consider two scenarios, hereinafter referred to as 2-link and 3-link scenario, respectively.
In the 2-link scenario, we consider a 10-MHz bandwidth LTE and an IEEE 802.11p link, and 5 MTs connected; the traffic load at the vRPA for each MT is  equal to 1\,Mbps. In the 3-link scenario, instead, we add a 5-MHz LTE link and consider 3 MTs, each associated  with 3-Mbps traffic load, plus 4 MTs, each associated with 1-Mbps traffic load. Fig.\,\ref{fig:context_vs_time} shows an instance of the time evolution of two context components, the SNR and the buffer state, with the latter referring to the per-MT 3\,Mbps downlink traffic load. Here we observe that the SNR is an independent variable and randomly takes values between $8$\,dB and $21$\,dB, while the evolution of the buffer state is action dependent, in the sense that over a course of time, as the algorithm is expected to learn to select better actions, the buffer state gradually reduces to zero. 
Finally, looking at the variation of the KPI values observed in our testbed  and whether they meet their respective thresholds, we set such thresholds 
as per the 3GPP specifications for 5G \cite{3gpp}, i.e., at $0.1$\,s for latency and 0.01 for packet loss.

\subsection{Convergence analysis}

\begin{figure*}[!t]
\centering
\subfloat[]{\includegraphics[width=0.333\linewidth]{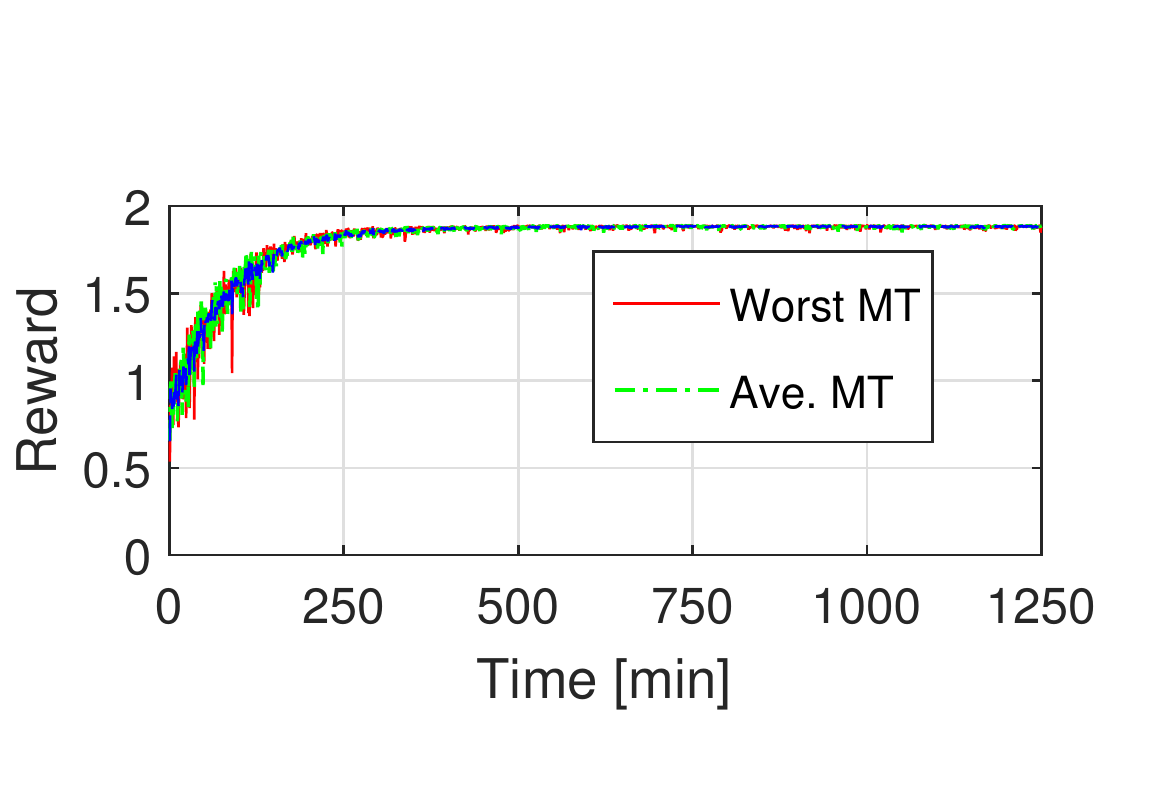}}
\subfloat[]{\includegraphics[width=0.333\linewidth]{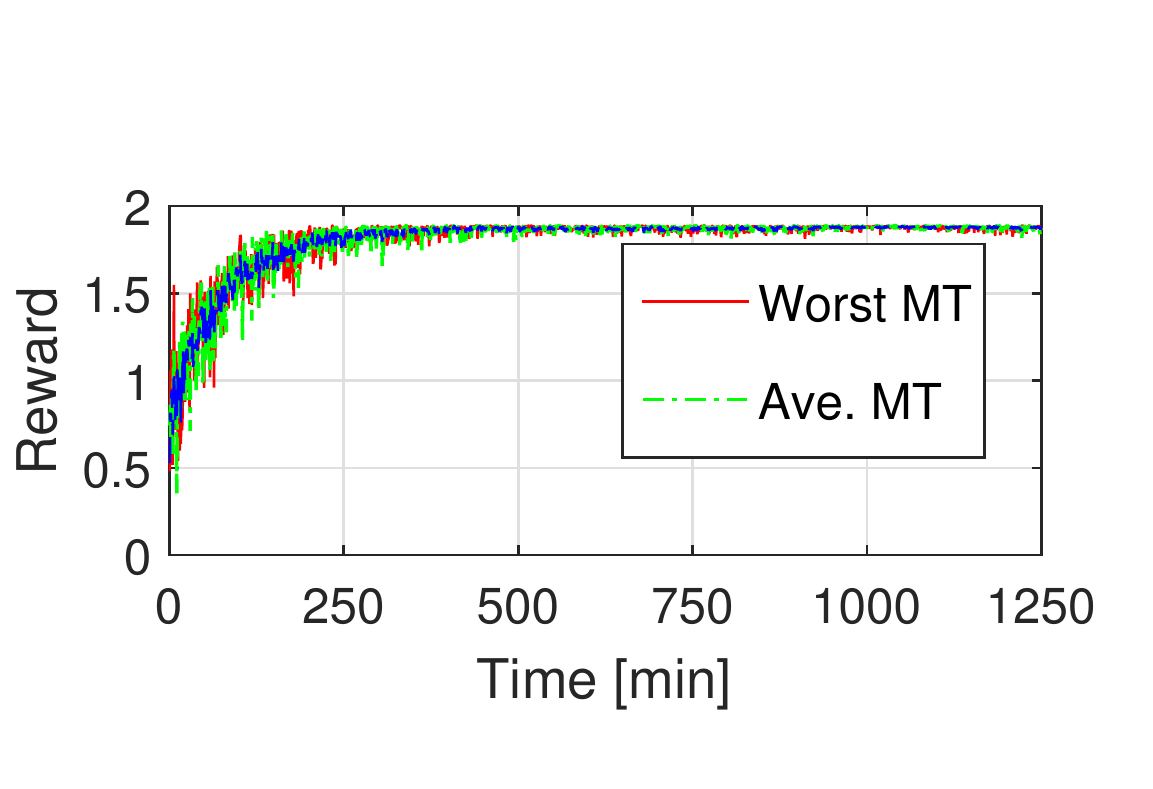}} 
\subfloat[]{\includegraphics[width=0.333\linewidth]{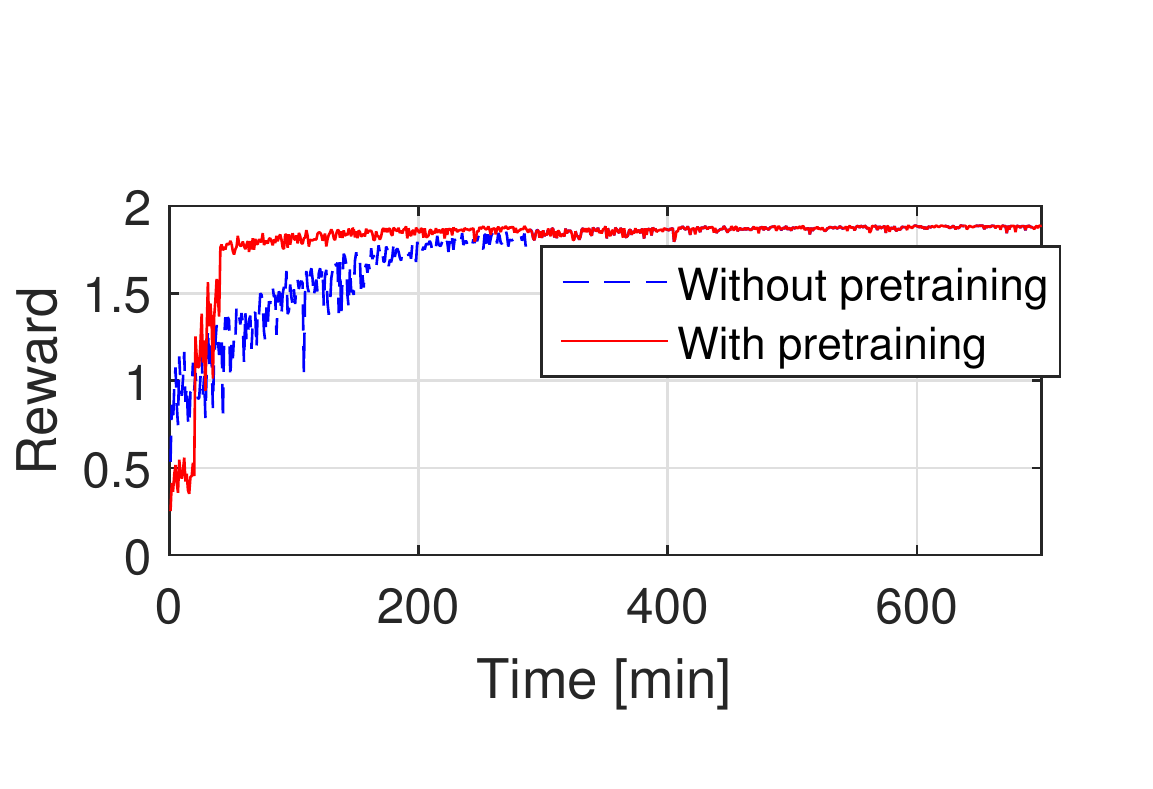}}
   \caption{Convergence of reward for decision making periodicity $N = 1$ (a) and $N = 10$ (b), with  reward corresponding to the  best (blue), worst (red), and average (green) MT performance. Comparison between CAREM convergence with and without pre-training, for $N = 1$ and worst MT (c). {Uniform convergence across different MTs and for different decision periodicities indicating efficient learning. Pre-training helps to achieve faster convergence.
    }}
\label{fig:rew_conv}
  \end{figure*}

We first focus on the 2-link scenario and evaluate the performance of \CaRReM\ in terms of convergence of reward values on time-sequenced context. 
The variation of reward values as a function of time  is depicted in Fig.\,\ref{fig:rew_conv}(a)-(b), for both $N=1$ and $N=10$ decision-making settings, and for  best, worst and average MT in the system. Here, the best (worst) MT is the one experiencing the highest (lowest) value of reward averaged over the experiment duration.  Instead, average MT is a benchmark scenario wherein during each monitoring slot the reward is evaluated by averaging the reward over all MTs. For both operational settings, we observe that the variation in the convergence behavior is negligible for best, worst, and average MT. Thus, we remark that even in the presence of multiple MTs, each having a different temporal evolution of context vector, the learning of the \CaRReM\ framework is efficient. Further, although the variation in reward values if higher for $N=10$, its convergence is as good as that for $N=1$. Thus, a longer decision making periodicity lowers the computation complexity with respect to per-slot decision making without affecting the convergence behavior of the algorithm.

Finally, Fig.\,\ref{fig:rew_conv}(c) shows how convergence can be further sped up when the RL model is pre-trained, as it is often deemed as required before a model starts operating in real-world systems. To highlight the difference between the cases with and without pre-training, we focus on a smaller time range on the plot x-axis and on the performance of the worst MT. It is  worth remarking that the limited difference between the two curves shown in the plot further  confirms that CAREM can quickly converge even in absence of pre-training.

\subsection{2-link scenario: KPIs, throughput, and action selection}
\begin{figure}[t!]
\centering
\subfloat[]{\includegraphics[width=\linewidth]{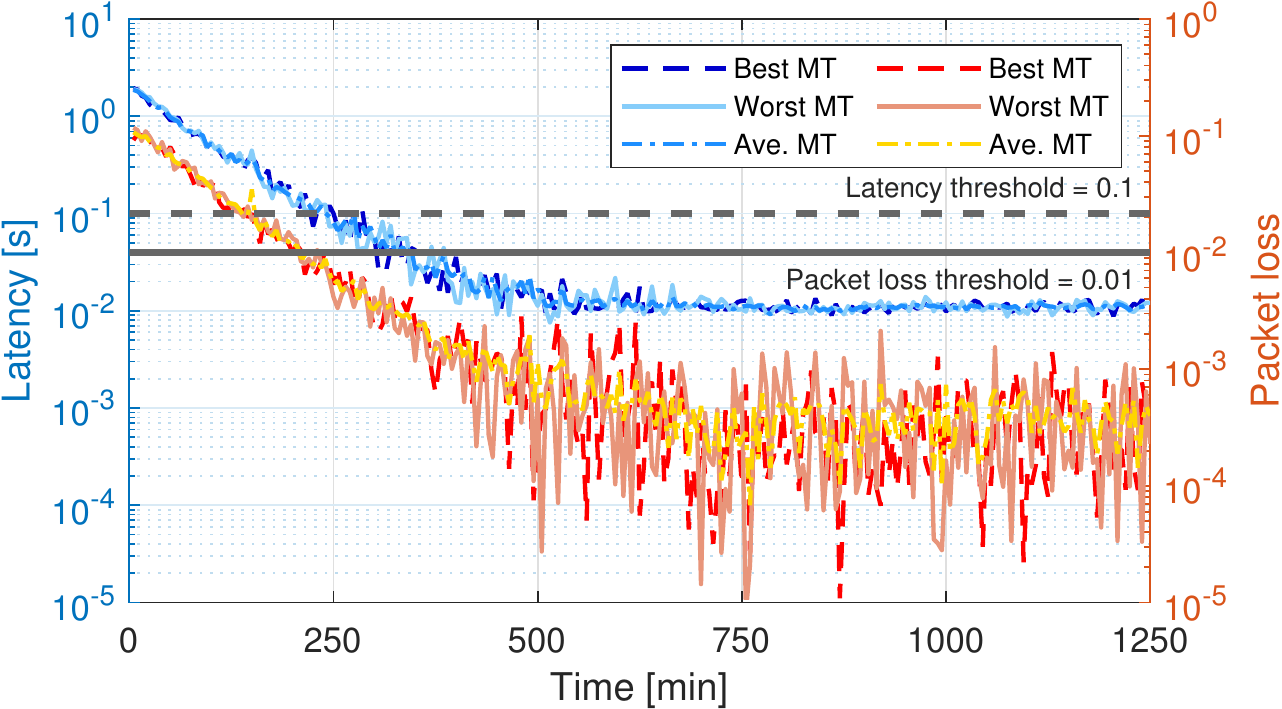}}\\
\subfloat[]{\includegraphics[width=\linewidth]{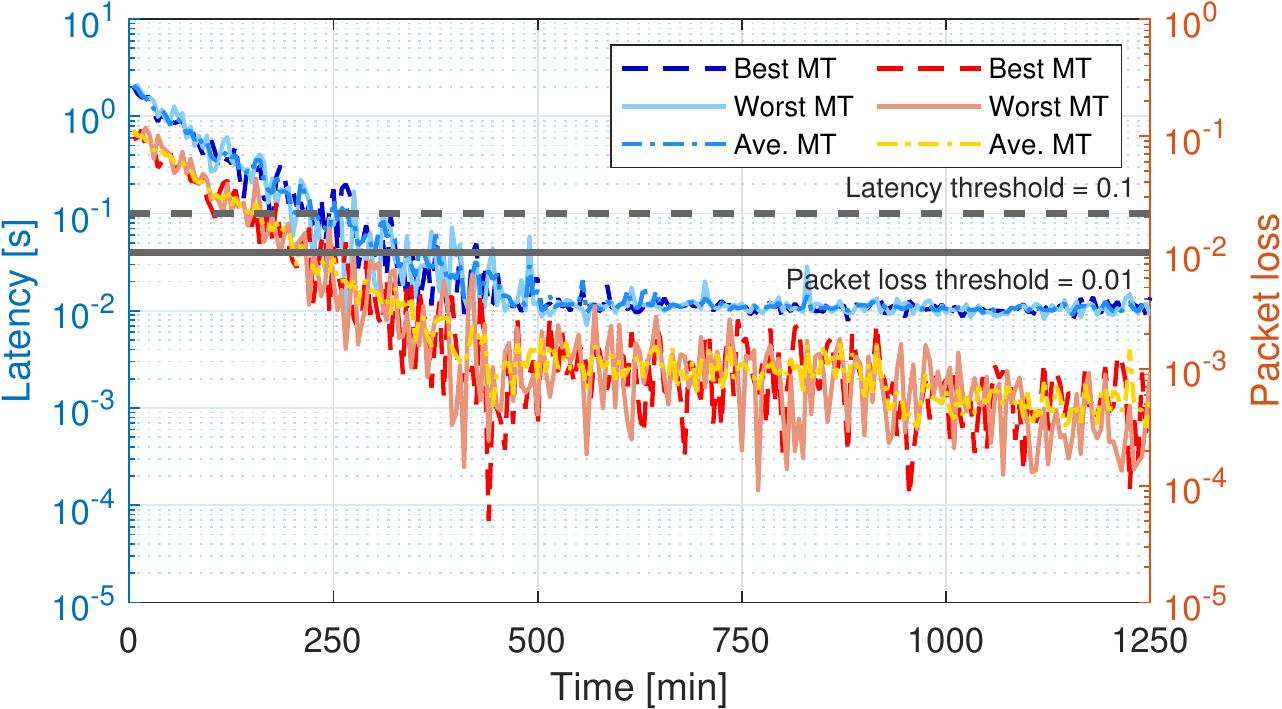}}
\caption{2-link scenario: Time evolution of the latency (blue shades) and packet loss (red shades) KPIs for the best (solid line),  worst (dashed line), and average (dashed dotted line) MT performance. KPI thresholds are depicted in gray.  
Decision making with periodicity $N = 1$ (a) and $N = 10$ (b).
{
KPI requirements  are met in all cases.} \label{fig:2L_kpi_v2_time}}
\vspace{-3mm}
\end{figure}

The results derived for the 2-link scenario, presented in  Fig.\,\ref{fig:2L_kpi_v2_time}, 
show that, except for an initial exploration period, the observed KPI values remain below their respective thresholds. This holds for all MTs, as can be seen by observing the curves referring to  the best,  worst and average MT performance. Compared to the $N=1$ decision making (Fig.\,\ref{fig:2L_kpi_v2_time}(a)), the observed packet loss is slightly higher for $N=10$ (Fig.\,\ref{fig:2L_kpi_v2_time}(b)), as in the latter case the action executed by  \CaRReM\  during a decision making interval may not be the optimum choice for all the slots in that interval. No significant degradation however is noticeable for either KPIs, thus suggesting that a larger decision making periodicity can be a viable solution to reduce the computational burden. 
Further, in case of pre-training (results omitted here for lack of room), CAREM can learn even faster, leading to a very significant reduction of the time during which KPIs are above threshold. 

\begin{figure}[t!]
\centering
\subfloat[]{\includegraphics[width=\linewidth]{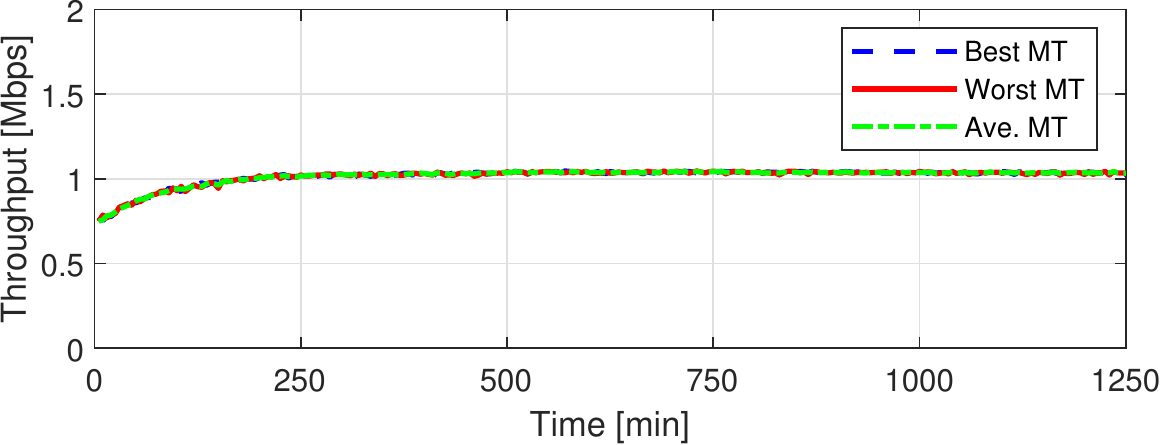}}\\
\subfloat[]{\includegraphics[width=\linewidth]{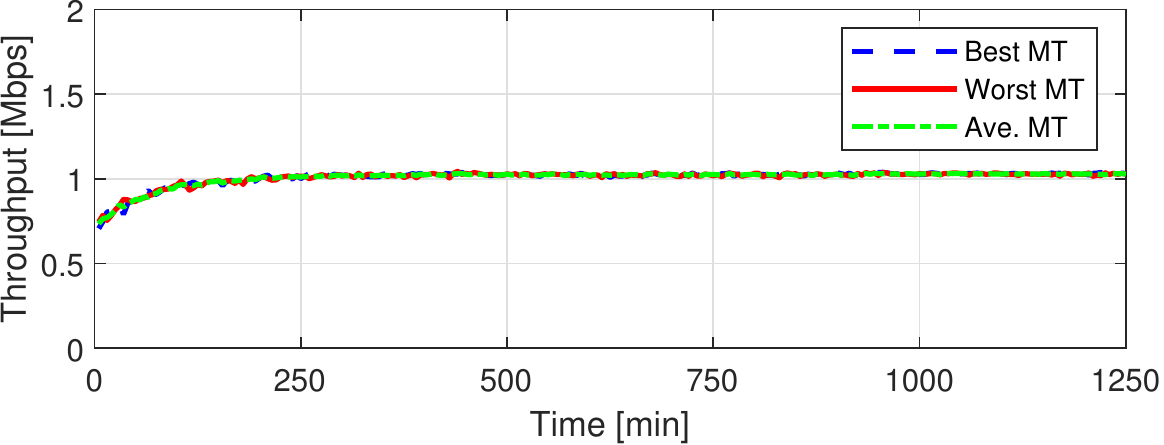}}
\caption{{2-link scenario: throughput in the case of best (blue),  worst (red), and average (green) MT performance, for $N = 1$ (a) and  $N = 10$ (b). For all the MTs, the measured throughput matches the offered  traffic load.}}
\label{fig:2L_throughput}
\vspace{-3mm}
\end{figure}

Next, we look at the throughput corresponding to the best, average, and worst MT performance. Even if throughput is not one of the KPIs targeted by CAREM, it is clearly correlated with latency and packet loss, and it is one of the reference metrics considered for the analysis of wireless systems.
As shown in Fig.\,\ref{fig:2L_throughput}, for all MTs the throughput matches the data traffic they are supposed to receive (i.e., 1\,Mbps),  thus confirming the effectiveness of CAREM. 

\begin{figure*}[t!]
\centering
\subfloat[]{\includegraphics[width=0.235\textwidth]{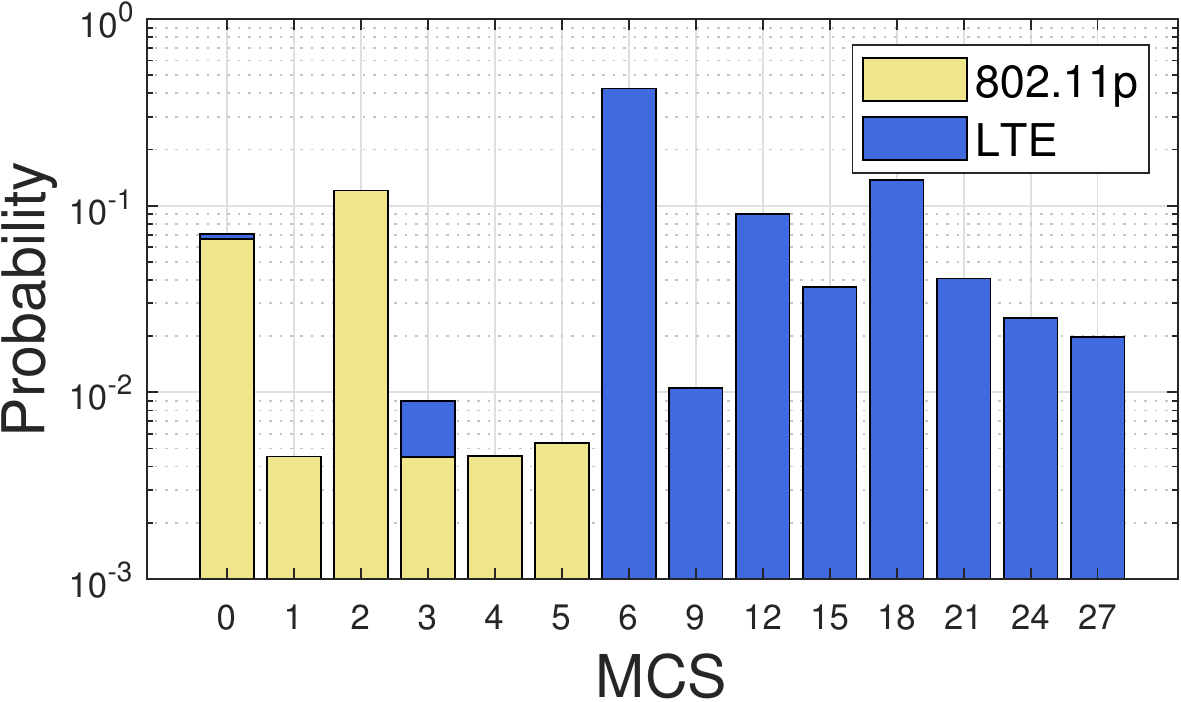}}
\hspace{-1.5mm}
\subfloat[]{\includegraphics[width=0.235\textwidth]{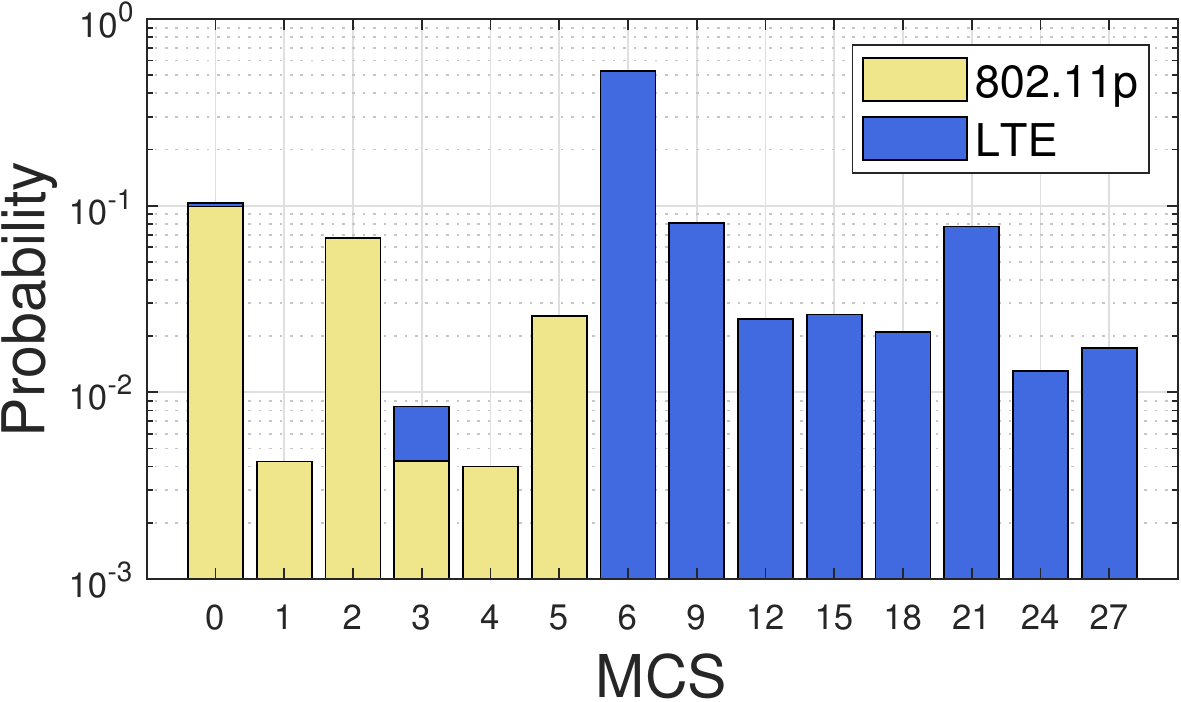}}
\hspace{-1.5mm} 
\subfloat[]{\includegraphics[width=0.235\textwidth]{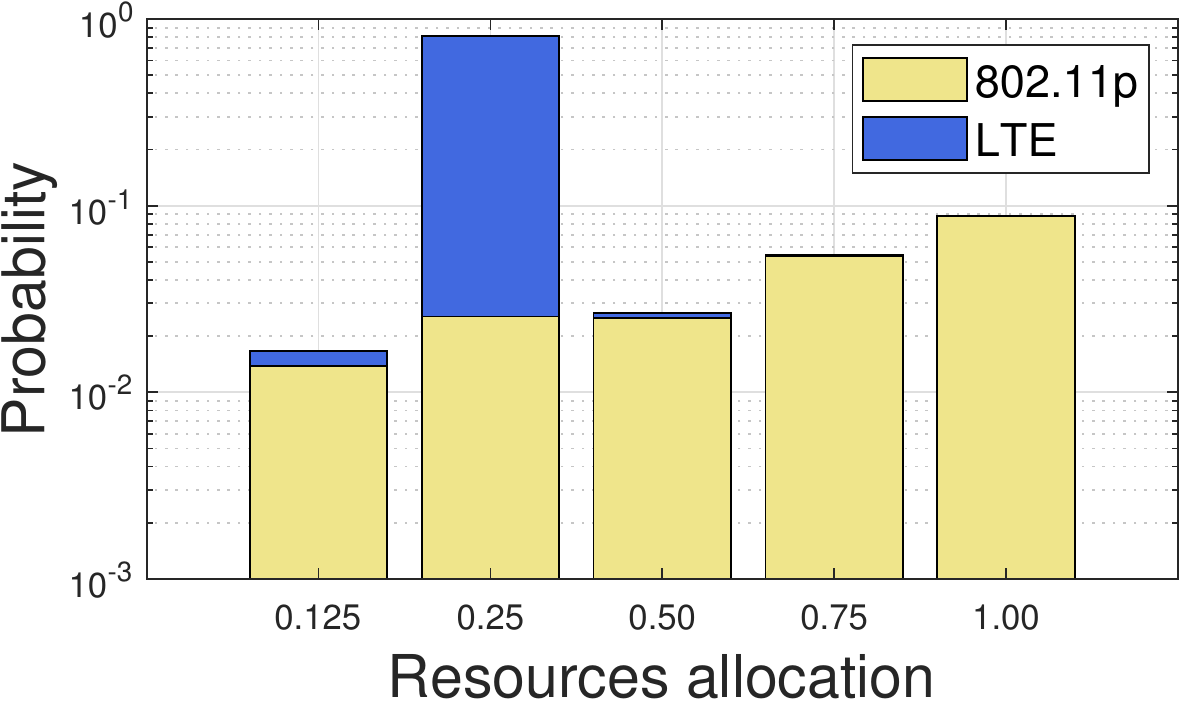}}
\hspace{-1.5mm}
\subfloat[]{\includegraphics[width=0.235\textwidth]{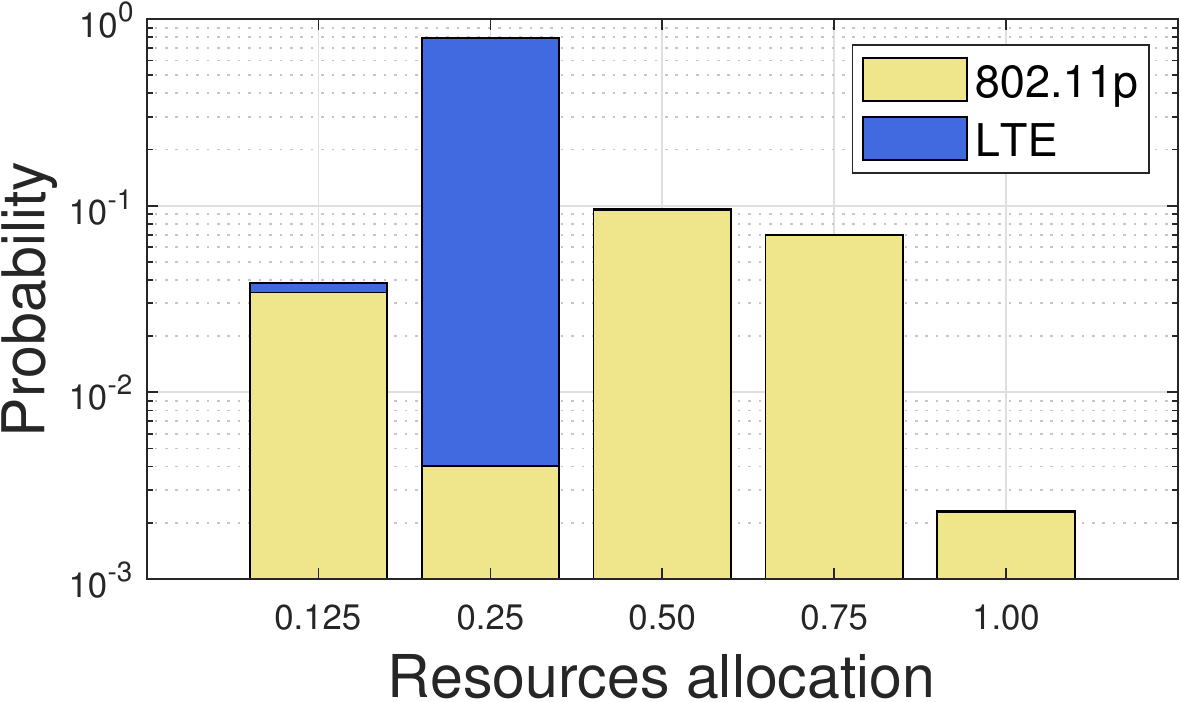}} 
\caption{2-link scenario: MCS selection for $N = 1$ (a) and $N = 10$ (b), 
and fraction of per-MT resource allocation for $N = 1$ 
{(c) and $N = 10$ (d). Bars in yellow and blue refer to IEEE 802.11p and 10-MHz LTE, 
respectively. The distribution of the selected MCSs ((a), (b)) indicates that CAREM can deal 
well with the correlation between MCS and the MTs’ state, while that of the link utilization  
((c), (d)) highlights that LTE is the most selected link.}} 
\label{fig:2L_mcs-res}
\vspace{-3mm}
\end{figure*}

Finally, Fig.\,\ref{fig:2L_mcs-res} depicts the frequency with which CAREM selects the values of MCS  (Fig.\,\ref{fig:2L_mcs-res}(a) and Fig.\,\ref{fig:2L_mcs-res}(b) for $N=1$ and $N=10$, resp.)  and resource allocation (Fig.\,\ref{fig:2L_mcs-res}(c) and Fig.\,\ref{fig:2L_mcs-res}(d) for $N=1$ and $N=10$, resp.). We remark that, with the aim to combine results for different capacity-constrained links,  the resource allocation fraction on the plots x-axis  is expressed as the ratio of   resources allocated  for each MT on a given link, to the number of available radio resources  thereof.

By looking at the results obtained for $N=1$ (Fig.\,\ref{fig:2L_mcs-res}(a) and (c)), one can observe that the choice of the MCS value varies depending on the experienced SNR as well as the link, as IEEE 802.11p supports only a subset of the MCS values that can instead be selected in LTE. On the latter link, it is quite evident a preference for relatively higher MCSs (greater or equal to 6), since, given a number of allocated RBs, such values allow for  higher throughput. 

With regard to link utilization in Fig.\,\ref{fig:2L_mcs-res}(c), the results reflect what the intuition suggests: LTE is the most used link. Indeed, since the 10-MHz LTE link  offers a higher capacity than  IEEE 802.11p, the latter  likely accommodates the traffic for only one MT at the time, while  LTE  is used for multiple MTs.  
Also, while the allocated air time on IEEE 802.11p varies depending on the number of packets to be transmitted to the MT using that link and the adopted MCS, the most likely number of RBs allocated on LTE for each accommodated traffic flow is equal to  0.24 (i.e., 12 RBs). Indeed, with the aim to best meet the target KPIs, CAREM tries to allocate as many resources as possible for the served MTs.  

At last, looking at  Fig.\,\ref{fig:2L_mcs-res}(b) and (d), which refer to $N=10$, we note that, although in this case the actions executed by \CaRReM\ may not be the optimum choice for all slots in a decision interval, its average resource utilization is almost at par with $N=1$. This is  further confirmed by the fact that the average resource utilization is found to be $14.91\%$ and $19.86\%$ for IEEE 802.11p and LTE (resp.) when $N=1$, and $11\%$ and $19.87\%$ for IEEE 802.11p and LTE (resp.) when $N=10$.


\begin{figure}[!t]
\centering{{\includegraphics[width=\linewidth]{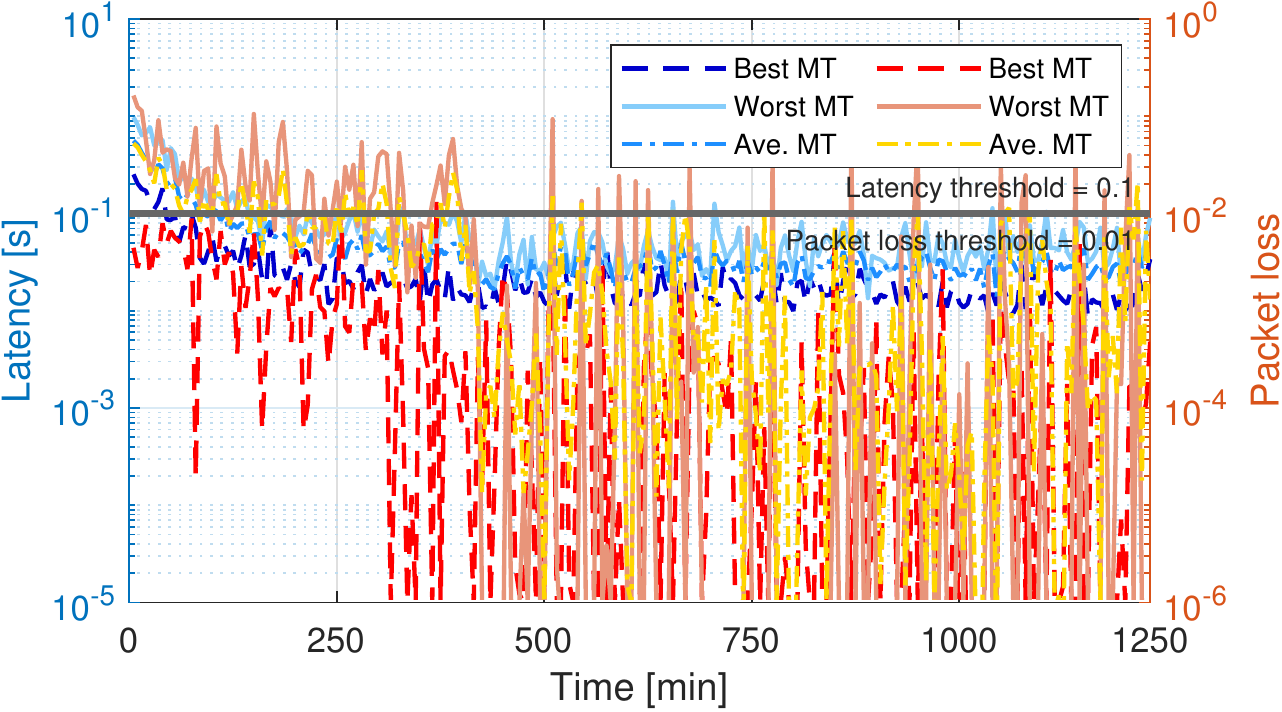} }}
   \caption{3-link scenario  for $N = 1$. KPIs time evolution (latency in shades of blue and packet loss in shades of red) 
for the best (solid line),  worst (dashed line), and average (dashed dotted line) MT performance (KPI thresholds are in gray). As the learning converges, average KPI thresholds are always met.}
\label{fig:3L_KPIevol}
\vspace{-3mm}
  \end{figure}
  
\begin{figure}[t!]
\centering
\subfloat[]
{\includegraphics[width=0.7\linewidth]{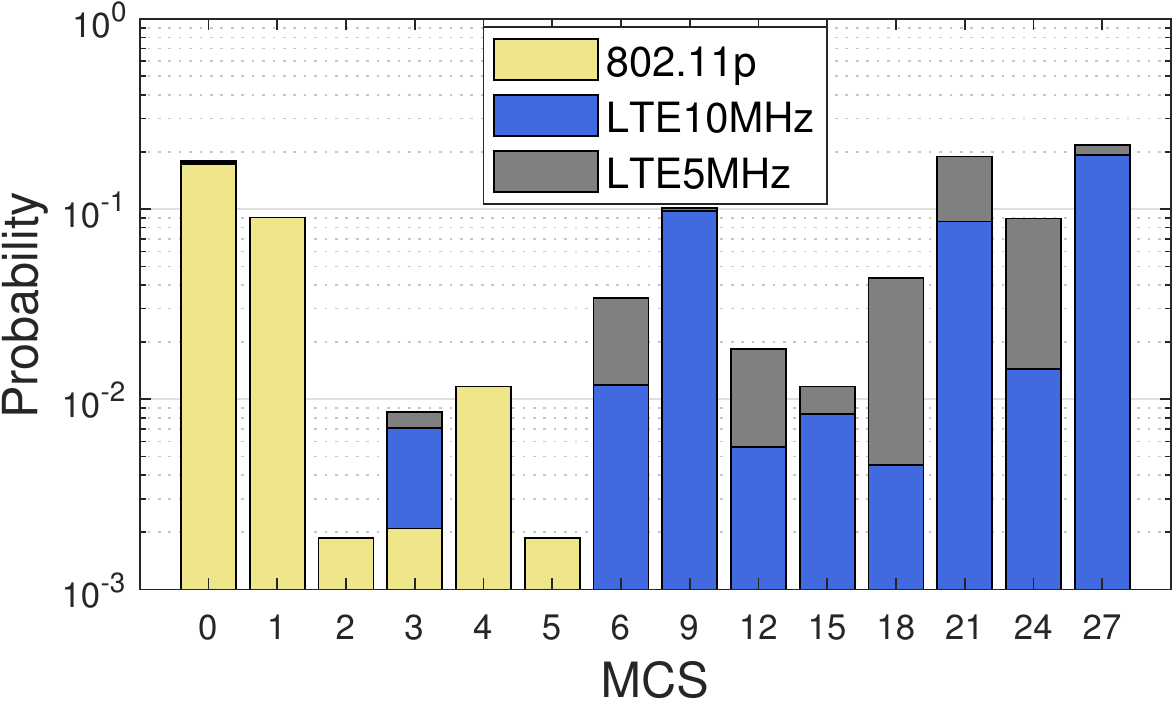}}\\
\subfloat[]
{\includegraphics[width=0.7\linewidth]{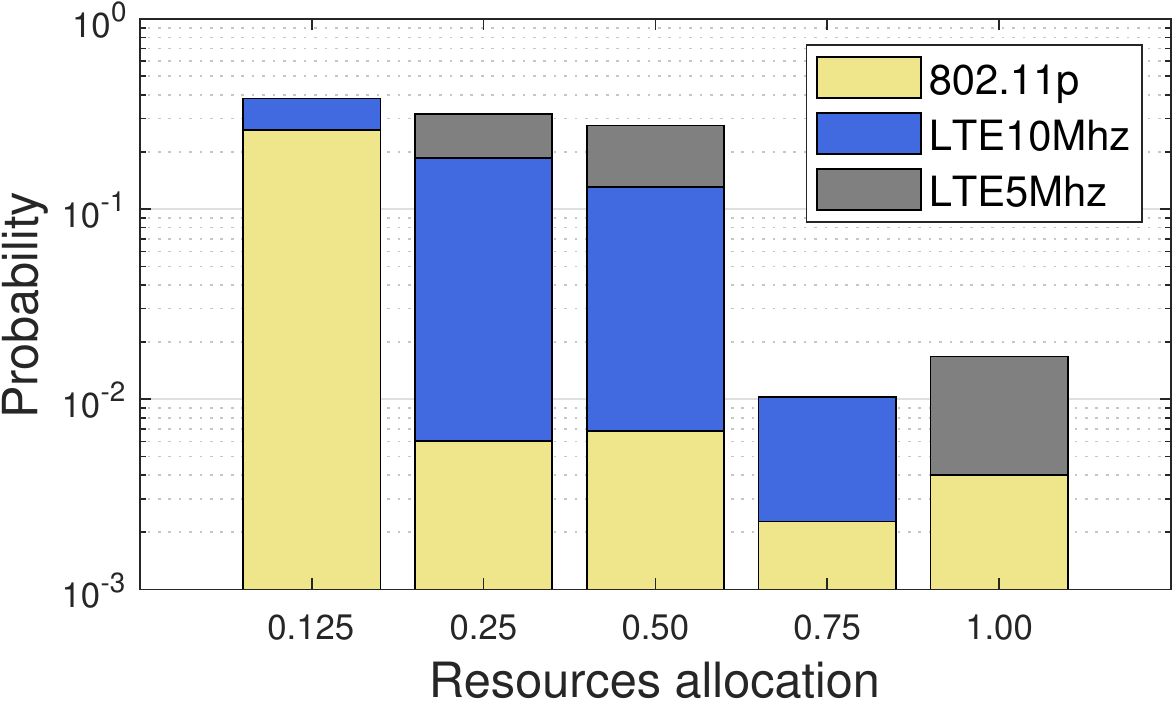}}
\caption{3-link scenario  for $N = 1$. Probability of selecting  MCS  (a) and resource allocation fraction (b).  In (a) and (b), bars in yellow, blue and gray refer to IEEE 802.11p, 10-MHz LTE, and 5-MHz LTE, respectively. Trends concur with 2-link scenario.}
\label{fig:3L_all}
\vspace{-3mm}
\end{figure}

\subsection{3-link scenario: KPIs and action selection}

To further show the scalability of CAREM, we now focus on the 3-link scenario and present the results in Fig.\,\ref{fig:3L_all} for $N=1$.
In particular, the plot in Fig.\,\ref{fig:3L_KPIevol} shows the latency (in shades of blue) and packet loss (in shades of red) for the best, worst and average MT performance.
After the initial learning, both KPIs  meet the target value,  except for the packet loss value of the MT experiencing the worst performance, which  exceeds the threshold in very few time instants. 

Fig.\,\ref{fig:3L_all}(a)--(b) depict instead the probability with which the different actions are selected, referring to the MCS values and the per-MT fraction of radio resource allocation  (resp.) on the three available links. Most of the trends observed for the 2-link case are confirmed. 
For 5-MHz LTE, we notice that  MCS values smaller than 27 are mostly selected, as the traffic load allocated on that link is lower than on 10-MHz LTE, hence less efficient, yet more robust, schemes are preferred. 
As for resource allocation  on the 5-MHz LTE link, we notice that the unitary value (i.e., 25 RBs) is selected with significantly higher probability, given the smaller link capacity.

\begin{figure}[t!]
\centering
\subfloat[]
{\includegraphics[width=\linewidth]{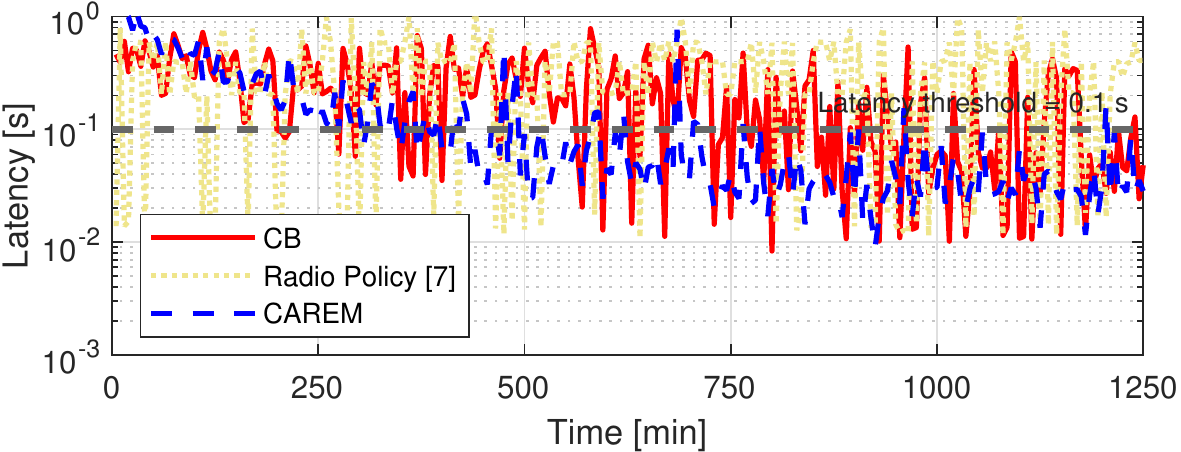}}\\
\subfloat[]
{\includegraphics[width=\linewidth]{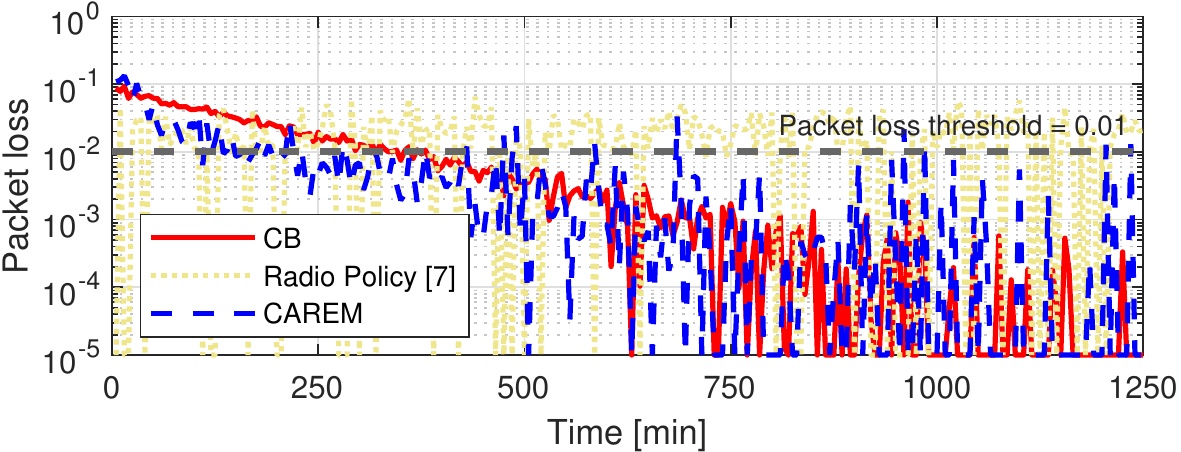}}
\caption{KPIs time evolution for CAREM, radio policy \cite{romero}, and contextual bandit (CB), in a one LTE-link scenario: latency (a) and packet loss (b).
{CAREM only can fulfill latency requirements, while target packet loss is met by CAREM and CB.}}
\label{fig:comp_KPI_CB_7}
\vspace{-3mm}
\end{figure}

\subsection{Comparative performance analysis}

Here, we compare the performance of \CaRReM\  with the closest competitive approach in \cite{romero}, a relatively simpler contextual bandit (CB) approach \cite{www2010_CB}, and standard LTE. For the latter, we rely on the srsRAN implementation,  which is compliant with LTE Rel.\,9. As for \cite{romero} (see also Sec.\,\ref{related_works}), after relaxing the computation constraints, the radio policy  targets the selection of suitable MCS for fast and reliable packet transmission over an LTE link. Note that  heterogeneous links are considered neither in \cite{romero} nor in LTE.  Hence, for the sake of fair comparison,  we  focus on the MCS allocation problem over a 10-MHz LTE link, and consider two MTs, each receiving a 3-Mbps traffic flow. 
In both \cite{romero} and CB, the classifier is trained using the dataset obtained from our testbed implementation of the LTE vRAN, as discussed in Sec.\,\ref{testbed_design}. Finally, in CB the same setting as in CAREM is used for the $\epsilon$ decay.

\begin{figure}[tb!]
\centering
\subfloat[]
{\includegraphics[width=0.8\linewidth]{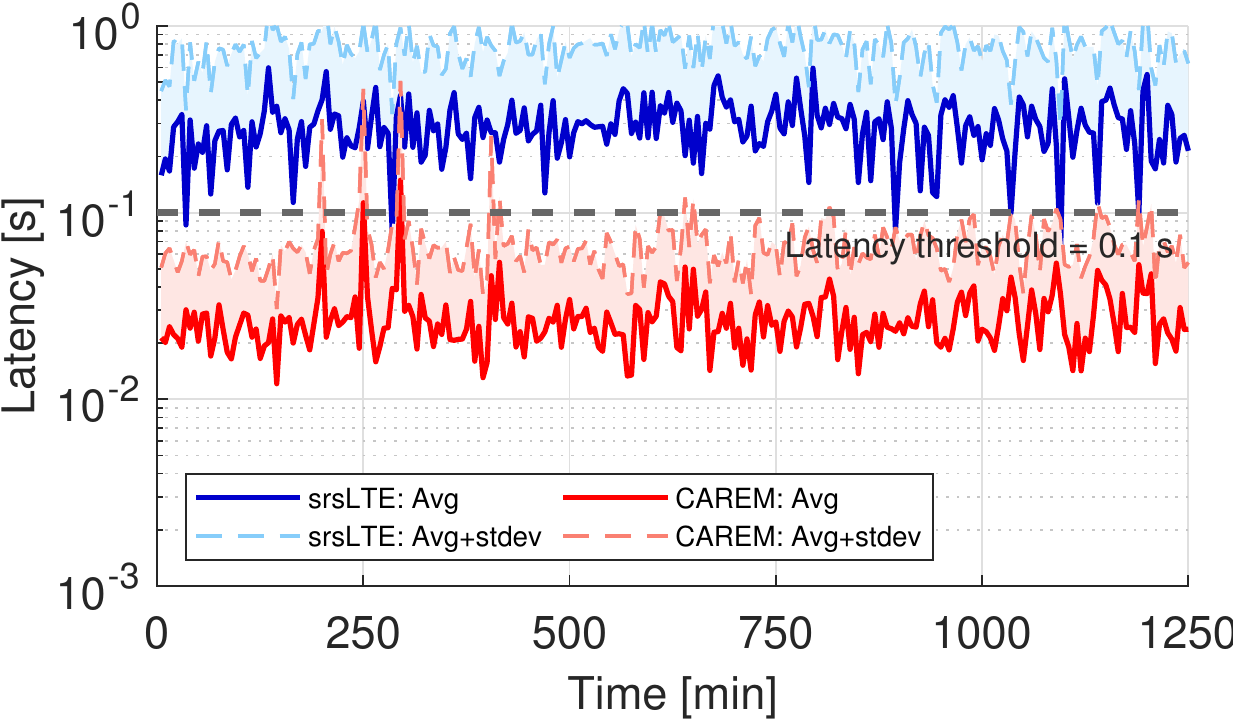}}\\
\subfloat[]
{\includegraphics[width=0.8\linewidth]{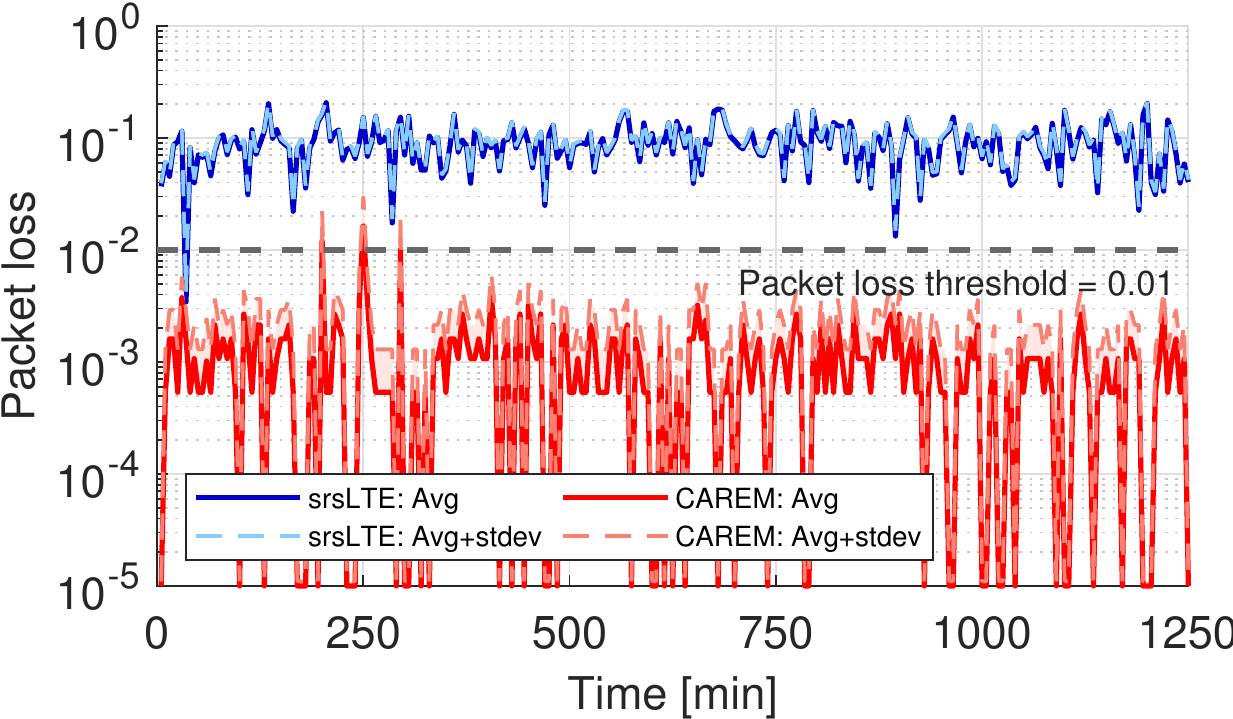}}
\caption{{KPIs time evolution for  one LTE link, under CAREM (red lines) and LTE (blue  lines).  Latency (a) and packet loss (b) averaged over two MTs (solid lines), and average plus standard deviation (dashed lines). CAREM outperforms LTE under high traffic load, owing to a more extensive exploration of different MCS values.}}
\label{fig:comp_srsLTE}
\vspace{-3mm}
\end{figure}

Fig.\,\ref{fig:comp_KPI_CB_7} presents the comparison of the KPI variation over time for $N=1$ for CB, the radio policy in \cite{romero} and  CAREM. We observe that, under CAREM and after the initial learning time, the variation of the latency in Fig.\,\ref{fig:comp_KPI_CB_7}(a) is essentially always below threshold. 
 The target latency value instead cannot be met using the MCS selection from the CB scheme or the radio policy in \cite{romero}: on average, CAREM provides a 65\% improvement with respect to CB and about one order of magnitude relatively to \cite{romero}. In case of CB, this is attributed to the fact that being a relatively simpler scheme, it is unable to capture the associative aspect of context-action mapping when the choice of current action influences the future values of the context. Further, the radio policy in \cite{romero} is trained using a loss function that minimizes the decoding error probability of the packet without accounting for latency as a criteria, consequently leading to larger latency values. 
This confirms that, thanks to the ability of SARSA to learn the best state-action trajectories, CAREM avoids taking high risk actions and incurring undesirable network states, which makes it a better match for dynamic scenarios than, e.g., a CB approach.

Further, from the packet loss variation in Fig.\,\ref{fig:comp_KPI_CB_7}(b), 
we note that  \CaRReM\ and CB provide similar performance, which is almost one order of magnitude  better than that of the radio policy in \cite{romero}. Indeed, the latter aims at limiting the bit error  rate at the physical layer, which may lead to different actions with respect to such policies as CAREM  targeting instead a desired level of packet loss at the MAC layer.

At last, Fig.\,\ref{fig:comp_srsLTE} compares the performance of CAREM against LTE, as implemented in srsRAN, again with two MTs and 3\,Mbps traffic load. For clarity of the plots, we depict   the value of the KPIs averaged over the MTs, along with the average value plus standard deviation. Note that, given the considered SNR pattern, the offered traffic load exceeds the maximum throughput that the standard LTE link can provide. Under such conditions, CAREM can instead support the considered traffic load with the target latency (Fig.\,\ref{fig:comp_srsLTE}(a)) and packet loss (Fig.\,\ref{fig:comp_srsLTE}(b)) values, although it has been designed primarily to manage heterogeneous links in vRANs. By looking at the  physical layer metrics (omitted here for lack of room), we noticed that the better  performance of CAREM is due to its ability of exploring various actions for a given context, hence more possible values of MCS. Some of them  are never selected by standard LTE, but they can actually lead to a higher reward and are therefore chosen by CAREM.

\section{Conclusions} \label{conclusion}
We have proposed \CaRReM, a novel RL based framework that efficiently allocates radio resources in terms of link, MCS, RBs and airtime for packet transmissions in heterogeneous vRANs. The choice of the RL algorithm, actions, and reward function has been made so that the resource utilization is optimized with respect to dynamic and non-stationary environments, with limited  computation efforts. Importantly, we have provided a proof-of-concept of our solution, by developing a testbed that leverages an LTE and an IEEE 802.11p SDR implementation. 
We have evaluated \CaRReM\ under different operational settings, with  different decision-making periodicity, number of links, number of MTs connected, and traffic load. 
The results show that, as the learning process of the model saturates, actions are chosen so that both  the observed KPIs, latency and packet loss, always satisfy their  target values. Further, it outperforms state-of-the art solutions as well as standard LTE, as implemented in the srsRAN framework. In comparison to the closest competitive scheme in \cite{romero} and LTE,  both the latency and packet loss observed with \CaRReM\  are of about one  order of magnitude lower, while CAREM provides a 65\% latency improvement relatively to contextual bandit. 
Finally, we remark that \CaRReM\ is a promising starting point to the development of heterogeneous networks, where the advantages of different radio technologies can be fully exploited to maximize the performance and the robustness of the network. Additionally, it effectively addresses the need for a solution that can swiftly  adapt to the underlying channel-network dynamics for context-aware radio resource allocation in heterogeneous vRANs.

\bibliographystyle{IEEEtran}
\bibliography{bib_rrl}

\begin{thebibliography}{10}
\providecommand{\url}[1]{#1}
\csname url@samestyle\endcsname
\providecommand{\newblock}{\relax}
\providecommand{\bibinfo}[2]{#2}
\providecommand{\BIBentrySTDinterwordspacing}{\spaceskip=0pt\relax}
\providecommand{\BIBentryALTinterwordstretchfactor}{4}
\providecommand{\BIBentryALTinterwordspacing}{\spaceskip=\fontdimen2\font plus
\BIBentryALTinterwordstretchfactor\fontdimen3\font minus
  \fontdimen4\font\relax}
\providecommand{\BIBforeignlanguage}[2]{{%
\expandafter\ifx\csname l@#1\endcsname\relax
\typeout{** WARNING: IEEEtran.bst: No hyphenation pattern has been}%
\typeout{** loaded for the language `#1'. Using the pattern for}%
\typeout{** the default language instead.}%
\else
\language=\csname l@#1\endcsname
\fi
#2}}
\providecommand{\BIBdecl}{\relax}
\BIBdecl

\bibitem{andrews}
J.~G. {Andrews} and et~al., ``What will {5G} be?'' \emph{IEEE J. Sel. Areas
  Commun.}, vol.~32, no.~6, pp. 1065--1082, June 2014.

\bibitem{wang}
C.~{Wang} and et~al., ``Cellular architecture and key technologies for {5G}
  wireless communication networks,'' \emph{IEEE Commun. Mag.}, vol.~52, no.~2,
  pp. 122--130, Feb. 2014.

\bibitem{olwal}
T.~O. {Olwal}, K.~{Djouani}, and A.~M. {Kurien}, ``A survey of resource
  management toward 5{G} radio access networks,'' \emph{IEEE Commun. Surveys
  Tuts.}, vol.~18, no.~3, pp. 1656--1686, 2016.

\bibitem{liang2015}
C.~{Liang} and F.~R. {Yu}, ``Wireless network virtualization: A survey, some
  research issues and challenges,'' \emph{IEEE Commun. Surveys Tuts.}, vol.~17,
  no.~1, pp. 358--380, 2015.

\bibitem{tsagkaris2015}
K.~{Tsagkaris}, G.~{Poulios}, P.~{Demestichas}, A.~{Tall}, Z.~{Altman}, and
  C.~{Destré}, ``An open framework for programmable, self-managed radio access
  networks,'' \emph{IEEE Commun. Mag.}, vol.~53, no.~7, pp. 154--161, 2015.

\bibitem{romero}
J.~A. Ayala-Romero, A.~Garcia-Saavedra, M.~Gramaglia, X.~Costa-Perez,
  A.~Banchs, and J.~J. Alcaraz, ``{VrAIn:} a deep learning approach tailoring
  computing and radio resources in virtualized {RANs},'' in \emph{ACM MobiCom},
  New York, NY, USA, 2019.

\bibitem{5G}
Ericsson, ``{5{G} Radio Access},'' \emph{Ericsson Rev.}, vol.~6, pp. 1--8, June
  2014.

\bibitem{fu2018}
Y.~{Fu}, S.~{Wang}, C.~{Wang}, X.~{Hong}, and S.~{McLaughlin}, ``Artificial
  intelligence to manage network traffic of 5{G} wireless networks,''
  \emph{IEEE Netw.}, vol.~32, no.~6, pp. 58--64, 2018.

\bibitem{hussain2020}
F.~{Hussain}, S.~A. {Hassan}, R.~{Hussain}, and E.~{Hossain}, ``Machine
  learning for resource management in cellular and {I}o{T} networks:
  Potentials, current solutions, and open challenges,'' \emph{IEEE Commun.
  Surveys Tuts.}, vol.~22, no.~2, pp. 1251--1275, 2020.

\bibitem{tang2020}
F.~{Tang}, Y.~{Kawamoto}, N.~{Kato}, and J.~{Liu}, ``Future intelligent and
  secure vehicular network toward 6{G}: Machine-learning approaches,''
  \emph{Proc. IEEE}, vol. 108, no.~2, pp. 292--307, 2020.

\bibitem{xiong2019}
Z.~{Xiong}, Y.~{Zhang}, D.~{Niyato}, R.~{Deng}, P.~{Wang}, and L.~{Wang},
  ``Deep reinforcement learning for mobile 5{G} and beyond: Fundamentals,
  applications, and challenges,'' \emph{IEEE Veh. Technol. Mag.}, vol.~14,
  no.~2, pp. 44--52, 2019.

\bibitem{zhengEbook}
R.~{Zheng} and C.~{Hua}, ``Sequential learning and decision-making in wireless
  resource management,'' \emph{Wireless Networks}, 2016.

\bibitem{3gpp}
``{3{GPP TS} 23.501 {V}16.3.0 {T}echnical Specification Group Services and
  System Aspects; { S}ystem Architecture for the 5{G} System (5{GS}); {S}tage
  2, ({R}elease 16) },'' 3{GPP}, 12 2019.

\bibitem{slivkins}
A.~Slivkins, ``Introduction to multi-armed bandits,'' \emph{Foundations and
  Trends in Machine Learning}, vol.~12, no. 1-2, pp. 1--286, 2019.

\bibitem{jang}
B.~{Jang}, M.~{Kim}, G.~{Harerimana}, and J.~W. {Kim}, ``{Q}-learning
  algorithms: A comprehensive classification and applications,'' \emph{IEEE
  Access}, vol.~7, pp. 133\,653--133\,667, 2019.

\bibitem{combes2015}
R.~{Combes} and A.~{Proutiere}, ``{Dynamic Rate and Channel Selection in
  Cognitive Radio Systems},'' \emph{IEEE J. Sel. Areas Commun.}, vol.~33,
  no.~5, pp. 910--921, May 2015.

\bibitem{combes2019}
R.~{Combes}, J.~{Ok}, A.~{Proutiere}, D.~{Yun}, and Y.~{Yi}, ``{Optimal Rate
  Sampling in 802.11 Systems: Theory, Design, and Implementation},'' \emph{IEEE
  Trans. Mobile Comput.}, vol.~18, no.~5, pp. 1145--1158, May 2019.

\bibitem{gupta2018}
H.~{Gupta}, A.~{Eryilmaz}, and R.~{Srikant}, ``{Low-complexity, Low-regret Link
  Rate Selection in Rapidly-varying Wireless Channels},'' in \emph{IEEE Conf.
  Comput. Commun.}, 2018, pp. 540--548.

\bibitem{ma2019}
J.~{Ma}, T.~{Nagatsuma}, S.~{Kim}, and M.~{Hasegawa}, ``A
  machine-learning-based channel assignment algorithm for {IoT},'' in
  \emph{ICAIIC}, 2019, pp. 1--6.

\bibitem{hasegawa2020}
S.~{Hasegawa}, S.~{Kim}, Y.~{Shoji}, and M.~{Hasegawa}, ``Performance
  evaluation of machine learning based channel selection algorithm implemented
  on {IoT} sensor devices in coexisting {I}o{T} networks,'' in \emph{IEEE
  CCNC}, 2020, pp. 1--5.

\bibitem{qureshi2020}
M.~A. {Qureshi} and C.~{Tekin}, ``Fast learning for dynamic resource allocation
  in {AI}-enabled radio networks,'' \emph{IEEE Trans. Cogn. Commun. Netw.},
  vol.~6, no.~1, pp. 95--110, 2020.

\bibitem{helou2015}
M.~{El Helou} and et~al., ``A network-assisted approach for {RAT} selection in
  heterogeneous cellular networks,'' \emph{IEEE J. Sel. Areas Commun.},
  vol.~33, no.~6, pp. 1055--1067, 2015.

\bibitem{nguyen2017}
D.~D. {Nguyen}, H.~X. {Nguyen}, and L.~B. {White}, ``Reinforcement learning
  with network-assisted feedback for heterogeneous {RAT} selection,''
  \emph{IEEE Trans. Wireless Commun.}, vol.~16, no.~9, pp. 6062--6076, 2017.

\bibitem{wei2018}
Y.~{Wei}, R.~Y. F.\, M.~{Song}, and Z.~{Han}, ``User scheduling and resource
  allocation in {HetNets} with hybrid energy supply: An actor-critic
  reinforcement learning approach,'' \emph{IEEE Trans. Wireless Commun.},
  vol.~17, no.~1, pp. 680--692, 2018.

\bibitem{morozs2015}
N.~{Morozs}, T.~{Clarke}, and D.~{Grace}, ``Heuristically accelerated
  reinforcement learning for dynamic secondary spectrum sharing,'' \emph{IEEE
  Access}, vol.~3, pp. 2771--2783, 2015.

\bibitem{raj2018}
V.~{Raj}, I.~{Dias}, T.~{Tholeti}, and S.~{Kalyani}, ``Spectrum access in
  cognitive radio using a two-stage reinforcement learning approach,''
  \emph{IEEE J. Sel. Topics Signal Process.}, vol.~12, no.~1, pp. 20--34, 2018.

\bibitem{comsa2018}
I.~{Comşa} and et~al., ``Towards 5{G}: A reinforcement learning-based
  scheduling solution for data traffic management,'' \emph{IEEE Trans. Netw.
  Service Manag.}, vol.~15, no.~4, pp. 1661--1675, 2018.

\bibitem{comsa2020}
I.~{Comșa}, R.~{Trestian}, G.~{Muntean}, and G.~{Ghinea}, ``{5MART}: A 5{G
  SMART} scheduling framework for optimizing {Q}o{S} through reinforcement
  learning,'' \emph{IEEE Trans. Netw. Service Manag.}, vol.~17, no.~2, pp.
  1110--1124, 2020.

\bibitem{zhou2020}
Y.~{Zhou}, F.~{Tang}, Y.~{Kawamoto}, and N.~{Kato}, ``Reinforcement
  learning-based radio resource control in 5{G} vehicular network,'' \emph{IEEE
  Wireless Commun. Lett.}, vol.~9, no.~5, pp. 611--614, 2020.

\bibitem{wang2018}
S.~{Wang}, H.~{Liu}, P.~H. {Gomes}, and B.~{Krishnamachari}, ``Deep
  reinforcement learning for dynamic multichannel access in wireless
  networks,'' \emph{IEEE Trans. Cogn. Commun. Netw.}, vol.~4, no.~2, pp.
  257--265, 2018.

\bibitem{naparstek2019}
O.~{Naparstek} and K.~{Cohen}, ``Deep multi-user reinforcement learning for
  distributed dynamic spectrum access,'' \emph{IEEE Trans. Wireless Commun.},
  vol.~18, no.~1, pp. 310--323, 2019.

\bibitem{zong2019}
C.~{Zhong}, Z.~{Lu}, M.~C. {Gursoy}, and S.~{Velipasalar}, ``A deep
  actor-critic reinforcement learning framework for dynamic multichannel
  access,'' \emph{IEEE Trans. Cogn. Commun. Netw.}, vol.~5, no.~4, pp.
  1125--1139, Nov. 2019.

\bibitem{zhang2019}
L.~{Zhang}, J.~{Tan}, Y.~{Liang}, G.~{Feng}, and D.~{Niyato}, ``Deep
  reinforcement learning-based modulation and coding scheme selection in
  cognitive heterogeneous networks,'' \emph{IEEE Trans. Wireless Commun.},
  vol.~18, no.~6, pp. 3281--3294, 2019.

\bibitem{li2018}
X.~{Li}, J.~{Fang}, W.~{Cheng}, H.~{Duan}, Z.~{Chen}, and H.~{Li},
  ``Intelligent power control for spectrum sharing in cognitive radios: A deep
  reinforcement learning approach,'' \emph{IEEE Access}, vol.~6, pp.
  25\,463--25\,473, 2018.

\bibitem{nasir2019}
Y.~S. {Nasir} and D.~{Guo}, ``Multi-agent deep reinforcement learning for
  dynamic power allocation in wireless networks,'' \emph{IEEE J. Sel. Areas
  Commun.}, vol.~37, no.~10, pp. 2239--2250, 2019.

\bibitem{gyawali2019}
S.~{Gyawali}, Y.~{Qian}, and R.~Q. {Hu}, ``Resource allocation in vehicular
  communications using graph and deep reinforcement learning,'' in \emph{IEEE
  GLOBECOM}, Dec. 2019, pp. 1--6.

\bibitem{chen2020}
X.~{Chen} and et~al., ``Age of information aware radio resource management in
  vehicular networks: A proactive deep reinforcement learning perspective,''
  \emph{IEEE Trans. Wireless Commun.}, vol.~19, no.~4, pp. 2268--2281, 2020.

\bibitem{ye2019}
H.~{Ye}, G.~Y. {Li}, and B.~F. {Juang}, ``Deep reinforcement learning based
  resource allocation for {V2V} communications,'' \emph{IEEE Trans. Veh.
  Technol.}, vol.~68, no.~4, pp. 3163--3173, 2019.

\bibitem{zhang2019_iot}
X.~{Zhang}, M.~{Peng}, S.~{Yan}, and Y.~{Sun}, ``Deep reinforcement learning
  based mode selection and resource allocation for cellular {V2X}
  communications,'' \emph{IEEE Internet Things J.}, pp. 1--1, Dec. 2019.

\bibitem{wons2021}
S.~{Tripath}, C.~{Puligheddu}, and C.~{Chiasserini}, ``An {RL} approach to
  radio resource management in heterogeneous virtual {RANs},'' in
  \emph{{IEEE/IFIP WONS}}, 2021.

\bibitem{5GT}
\BIBentryALTinterwordspacing
``5{G} mobile transport platform for verticals,'' \emph{5{G PPP H}2020
  5{G-TRANSFORMER} Project}, 2019. [Online]. Available:
  \url{http://5g-transformer.eu}
\BIBentrySTDinterwordspacing

\bibitem{5Growth}
\BIBentryALTinterwordspacing
``5{G}-enabled growth in vertical industries,'' \emph{5{G PPP H}2020 5Growth
  Project}, 2021. [Online]. Available: \url{https://5growth.eu}
\BIBentrySTDinterwordspacing

\bibitem{sutton}
R.~S. Sutton and A.~G. Barto, \emph{Reinforcement Learning: An
  Introduction}.\hskip 1em plus 0.5em minus 0.4em\relax Cambridge, MA, USA: MIT
  Press, 1998.

\bibitem{tile_coding}
A.~A. Sherstov and P.~Stone, ``Function approximation via tile coding:
  Automating parameter choice,'' in \emph{Abstraction, Reformulation and
  Approximation}, J.-D. Zucker and L.~Saitta, Eds.\hskip 1em plus 0.5em minus
  0.4em\relax Berlin, Heidelberg: Springer Berlin Heidelberg, 2005, pp.
  194--205.

\bibitem{laumanns2002}
M.~{Laumanns}, L.~{Thiele}, K.~{Deb}, and E.~{Zitzler}, ``Combining convergence
  and diversity in evolutionary multiobjective optimization,''
  \emph{Evolutionary Computation}, vol.~10, no.~3, pp. 263--282, 2002.

\bibitem{miguelez}
I.~Gomez-Miguelez and et~al., ``{SrsLTE:} an open-source platform for {LTE}
  evolution and experimentation,'' in \emph{ACM WiNTECH}, 2016, p. 25–32.

\bibitem{bloessl}
B.~Bloessl, M.~Segata, C.~Sommer, and F.~Dressler, ``Performance assessment of
  {IEEE} 802.11p with an open source {SDR}-based prototype,'' \emph{IEEE Trans.
  Mobile Comput.}, vol.~17, no.~5, pp. 1162--1175, May 2018.

\bibitem{www2010_CB}
L.~Li, W.~Chu, J.~Langford, and R.~Schapire, ``{A Contextual-bandit Approach to
  Personalized News Article Recommendation},'' in \emph{ACM WWW}, 2010.

\end{thebibliography}

\end{document}